\documentclass[preprint,a4paper,fleqn]{cas-dc}
\usepackage{cas-common}
\usepackage{graphicx}
\usepackage{float}
\usepackage{datetime}
\usepackage{amsmath,amsfonts,amssymb,mathrsfs}
\usepackage{natbib}
\usepackage{epsfig}
\usepackage{color}
\usepackage{bm}
\usepackage{booktabs}
\usepackage{array}
\usepackage{multirow}
\usepackage{placeins}
\usepackage{caption}
\usepackage{subfig}
\usepackage{enumitem}
\usepackage{afterpage}
\usepackage{etoolbox}

\DeclareUnicodeCharacter{202F}{FIX ME!!!!}

\def\tsc#1{\csdef{#1}{\textsc{\lowercase{#1}}\xspace}}
\tsc{WGM}
\tsc{QE}

\begin{document}

\let\WriteBookmarks\relax
\def\floatpagepagefraction{1}
\def\textpagefraction{.001}

\title[mode = title]{An open-source finite element toolbox for anisotropic creep and irradiation growth: Application to tube and spacer grid assembly
}

\shorttitle{An open-source finite element toolbox for anisotropic creep and irradiation growth: Application to tube and spacer grid assembly}    
\author[1]{Fabrizio E. Aguzzi}[]
\author[1]{Santiago M. Rabazzi}[]
\author[1]{Mart\'in S. Armoa}[]


\author[1]{César I. Pairetti}[]
\author[2]{Alejandro E. Albanesi}[]
\cormark[1]
\ead{aalbanesi@cimec.santafe-conicet.gov.ar}
\cortext[1]{Corresponding author}

\affiliation[1]{organization={IFIR Instituto de F\'isica de Rosario, CONICET-UNR},
	addressline={Ocampo y Esmeralda, Predio CONICET Rosario}, 
	city={Rosario},
	postcode={2000}, 
	state={Santa Fe},
	country={Argentina}}

\affiliation[2]{organization={CIMEC Centro de Investigaci\'on de M\'etodos Computacionales, CONICET-UNL},
            addressline={Col. Ruta 168 s/n, Predio CONICET Santa Fe}, 
            city={Santa Fe},
            postcode={3000}, 
            state={Santa Fe},
            country={Argentina}}

\begin{abstract}
This work presents an open-source interface that couples the viscoplastic self-consistent (VPSC) model—capable of simulating anisotropic creep and irradiation growth in polycrystalline materials—with the finite element solver Code\_Aster. The interface enables the simulation of the micromechanical response of irradiated zirconium alloy components by integrating grain-level constitutive behavior into a structural FEM framework. A key feature is the automated rotation of stress and strain tensors between the global FEM frame and the local crystallographic axes, a transformation not natively supported by Code\_Aster. The elastic strain is recovered analytically using the inverse of the self-consistently  stiffness tensor provided by VPSC.

As a demonstration, the framework is applied to an actual model of a pressurized water reactor (PWR) spacer grid, based on a patented design, capturing nonlinear contact and the anisotropic response of the cladding and grid. Simulations reveal the micromechanisms controlling the evolution of clearance between components and highlight the role of crystallographic texture in mitigating wear. In particular, a texture with a high fraction of prismatic planes oriented in the normal direction of the grid appears to be the most suitable for spacer design, as it minimizes clearance and contributes to wear resistance. The interface offers a flexible, extensible platform for high-fidelity simulations in nuclear fuel performance analysis.
\end{abstract}



\maketitle

\section{Introduction}\label{sec:intro}

The mechanical integrity of nuclear fuel cladding under irradiation and complex loading conditions is a critical factor in ensuring fuel performance and reactor safety. Conventional fuel performance codes frequently rely on empirical constitutive models calibrated for specific materials and service conditions. While effective within narrow operating ranges, these models lack predictive capability when applied to advanced cladding designs or evolving operational demands. In parallel, the adoption of open-source, validated simulation platforms is increasingly vital to promote transparent, reproducible, and extensible tools in nuclear engineering.

\begin{figure}[h!] 
	\centering
	\includegraphics[width=0.24\textwidth]{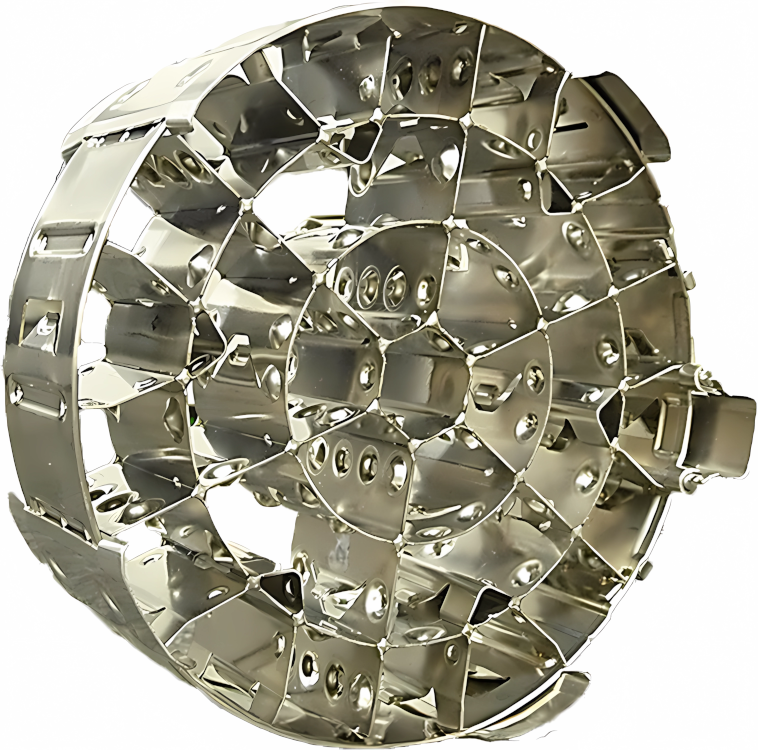}
	\caption{Flexible Spacer Grid \citep{CONUAR}.}  
	\label{fig:FlexibleSpacerGrid}
\end{figure}

In pressurized water reactors (PWRs), fuel rod assemblies consist of uranium dioxide pellets enclosed in zirconium alloy cladding tubes, held in place by spacer grids. These grids incorporate spring and dimple elements that apply a mechanical preload on the cladding, ensuring contact and alignment during operation (see Fig.~\ref{fig:FlexibleSpacerGrid}). Under irradiation, zirconium alloys undergo dimensional changes driven by creep and irradiation growth~\citep{fidleris1988irradiation}. These deformation mechanisms can lead to the gradual loss of contact between the cladding and the spacer grid elements. Combined with vibrations induced by the coolant flow, this loss of contact can initiate fretting wear, compromising the integrity of the cladding surface~\citep{kim2009impact, yan2011new, kim2008geometry, choi2004vibration}. As a result, wear-induced failures remain a key concern in the design and qualification of fuel rod assemblies~\citep{kim2010study}.

Early modeling efforts typically treated creep and irradiation growth as separate phenomena, using predefined assumptions to estimate gap evolution between the cladding and the grid. While these approaches captured the overall trends, they provided limited insight into the underlying micromechanical mechanisms. More recent works have introduced coupled models to simultaneously capture the interactions between irradiation growth and creep deformation~\citep{billerey2005evolution}, yet many of these remain phenomenological in nature. As a result, they are constrained in their ability to account for microstructural features and material history—such as crystallographic texture, prior mechanical working, or thermal processing.

Creep and irradiation growth are strongly dependent on the microstructure of zirconium alloys and have been widely investigated~\citep{fidleris1988irradiation, griffiths1995microstructural}. A significant advancement was proposed by \citet{patra2017crystal}, who developed a crystal plasticity framework for polycrystalline materials with a hexagonal close-packed (HCP) structure. This model incorporates the generation of irradiation defects and their interactions with dislocations and grain boundaries, including the nucleation and evolution of dislocation loops. The grain-level response is modeled through a single crystal description, and the overall polycrystalline behavior is obtained using the viscoplastic self-consistent (VPSC) formulation~\citep{tome1993self,tome2023material}, which captures the evolution of texture and accommodates both creep and irradiation effects. In this framework, a finite element method (FEM) solver is required to compute the mechanical equilibrium at the structural level.

In this context, the present work introduces an open-source interface, referred to as VPSC-CAFEM, that couples the micromechanical VPSC model with the finite element solver Code\_Aster~\cite{ASTER}. It enables the simulation of anisotropic, texture-dependent deformation of irradiated zirconium components by integrating grain-level constitutive behavior. Stress and strain tensors are computed using VPSC and mapped onto the material properties of Code\_Aster through a UMAT-like structure. The formulation is based on an additive decomposition of the strain increment, where the viscoplastic part is computed by the VPSC model and the elastic part is calculated analytically using the inverse of the self-consistent elastic stiffness tensor provided by VPSC. Although the open-access version of VPSC does not explicitly incorporate elasticity, this tensor allows the elastic response to be recovered analytically and incorporated into the FEM solution to ensure mechanical equilibrium ~\citep{llubel2024integracion}. This approach offers physics-based alternative to the semi-empirical models currently available in the standard version of Code\_Aster.

A key feature of the interface is its capability to automatically rotate stress and strain tensors between the global frame used in the finite element solver and the local crystallographic reference frame required by VPSC. This transformation, not supported by default in Code\_Aster, is carried out at each integration point, enabling large-scale simulations involving millions of Gauss points. The accuracy of the implementation is validated against the benchmark results reported in~\cite{patra2017crystal} for irradiated Zircaloy cladding. Furthermore, the framework is applied to an industrial-scale model of a PWR spacer grid with its current patented geometry \citep{CONUAR}, capturing the nonlinear contact between cladding and grid elements and the anisotropic response of the material—all within a fully open-source environment.

The modular structure of the developed interface facilitates future extensions to incorporate additional phenomena such as thermal expansion, thermal creep, rate-dependent plasticity, and precipitation hardening, among others~\citep{montgomery2017use}. Overall, this work contributes a flexible, reproducible, and physics-based simulation platform for high-fidelity structural analysis in nuclear fuel applications.

\FloatBarrier

\section{Numerical Model}\label{sec:case}
In the context of solving nonlinear quasi-static problems in structural mechanics, the FEM solver must address the following system of equations~\cite{agouzal2024reduced}:

\begin{equation}
\left\{
\begin{aligned}
    -\nabla\cdot\sigma &= f \\  
    \sigma &= F^{\sigma}(\nabla_s u,\beta) + \ (BCs)\\
    \dot{\beta} &= F^\beta(\sigma,\beta)
\end{aligned}
\right.
\end{equation}

\noindent where \(u\) is the displacement field, \(\sigma\) is the stress tensor field, \(\beta\) are the internal variables that characterize the microstructure, including the orientation, morphology, and hardening state of each statistically representative grain in this study. The nonlinear operator \(F^\sigma\) stands for the constitutive equation that maps the stress state in the material from the deformations state, the symmetric part of the gradient, \mbox{\(\nabla_s =\frac{1}{2}(\nabla + \nabla^T)\)}, while the nonlinear operator \(F^\beta\) denotes an evolution equation of internal variables within the material. The first equation of the system above describes the physical equilibrium on the considered domain. On the second line, the constitutive equation provides the stress state of the material, the Cauchy stress tensor  \(\sigma\), as a function of the deformation of the system (\(\nabla_s\ u\)) and the internal variables (\(\beta\)) of the material. The third equation describes the dynamics for the internal variables, giving a description of the material evolution. These variables play a crucial role in characterizing how the material responds to external influences over time.

\subsection{Micro-Scale Model}
\label{EcCristalSimple}

\subsubsection{Irradiation-induced growth crystallography model}

In this work we use the model reaction-diffusioin developed and implemented by ~\cite{patra2017crystal} .Therefore, the main equations describing the growth phenomenon under irradiation are presented.

To describe the deformation components and Burger vectors of the dislocations, four crystallographic directions are used: three compact basal directions, \( j \equiv \mathbf{a}_1, \mathbf{a}_2, \mathbf{a}_3 \), and one direction along the c-axis, \( j \equiv \mathbf{c} \). In addition the density of total dislocations (nucleating and growing in the prismatic and basal planes) can be calculated as the sum of line dislocations (with density $\rho_d$) plus intestitial-type dislocation loops (with number density \( N_{i}^{j} \) and mean radius \( r_{i}^{j} \)) plus vacancy type dislocation loops (with number density \( N_{v}^{j} \) and mean radius \( r_{v}^{j} \)).
\begin{equation}
\rho^{j} = \rho_d + 2 \pi r_{i}^{j} N_{i}^{j} + 2 \pi r_{v}^{j} N_{v}^{j};
\end{equation}

The climbing of dislocation loops and line dislocations during irradiation, contribute to positive growth strain along the prismatic directions, \(\varepsilon_{climb}^{j}\), and negative growth strain along the basal direction \(\varepsilon_{growth}^{c}\).
At steady state, the rate of crystallographic deformation along the prismatic and basal directions due to dislocation climbing, both line and dislocation loops, is given by the following equation:

\begin{equation} \label{epsilonClimb_a}
\begin{aligned}
\dot{\varepsilon}_{\text{climb}}^j &= G_{\text{NRT}} (1 - f_r) f_{ic} \left( \frac{1}{3} - \frac{\rho^j}{\sum_j \rho^j + \sum_j \sum_m k_{jm}^j} \right), \\
j &\equiv \mathbf{a}_1, \mathbf{a}_2, \mathbf{a}_3; \quad m \equiv \mathbf{x'}, \mathbf{y'}, \mathbf{z'}.
\end{aligned}
\end{equation}

\begin{equation}\label{epsilonClimb_c}
\begin{aligned}
\dot{\varepsilon}_{\text{climb}}^c &= -G_{\text{NRT}} (1 - f_r) f_{ic} \left( \frac{\rho^c}{\rho^c + \sum_m k_{cm}^c} \right);
\end{aligned}
\end{equation}

\noindent where the rate of defect production is the Norgett-Robinson-Torrens \(\ G_{\text{NRT}} \)~\cite{norgett1975proposed} for vacancies and interstitials. The fraction of point defects that recombine during the reaction is given by \(\ f_{\text{r}} \) and \(\ f_{\text{ic}} \) is the fraction of interstitials that form clusters. On the other hand, \( k_{m}^{j} \) is the grain boundaries (GBs) absorption strength of  for point defects (the super-index \( j \) indicates the direction of the Burger vector, while the sub-index \( m \) indicates the direction of one of the principal axes of the grain).

Therefore, the effect of defect absorption at grain boundaries makes an additional contribution to the growth strain:

\begin{equation}
\begin{aligned}  
\dot{\varepsilon}_{\text{GB}}^j &= -G_{\text{NRT}} (1 - f_r) f_{ic} \frac{\sum_m k_{m}^{j2} }{\sum_j \rho^j + \sum_j \sum_m k_{jm}^j} ; \\
\quad j &\equiv \mathbf{a}_1, \mathbf{a}_2, \mathbf{a}_3,\mathbf{c}; \quad m \equiv \mathbf{x'}, \mathbf{y'},\mathbf{z'}
\end{aligned}
\end{equation}

Finally, the rate of crystallographic deformation due to growth under irradiation, taking into account the above, is given by:

\begin{equation}
\dot{\varepsilon}_{\text{growth}}^j = \dot{\varepsilon}_{\text{climb}}^j + \dot{\varepsilon}_{\text{GB}}^j; j \equiv \mathbf{a}_1, \mathbf{a}_2, \mathbf{a}_3;\mathbf{c};
\end{equation}
\

The strain rate in the single crystal associated with growth under irradiation \(\dot{\varepsilon}^{(\text{growth})}\) is obtained by projecting the crystallographic strain rate \(\dot{\varepsilon}_{\text{growth}}^{j}\) along the principal axes of the crystal:

\begin{equation}
\label{EcuacionCrecimiento}
\begin{aligned}
\dot{\varepsilon}_{kl}^{(\text{growth})} &= \sum_j \dot{\varepsilon}_{\text{growth}}^j +b_{k}^j b_{l}^i; \\ j &\equiv \mathbf{a}_1, \mathbf{a}_2, \mathbf{a}_3;\mathbf{c}; k,l \equiv \mathbf{x}, \mathbf{y}, \mathbf{z};
\end{aligned}
\end{equation} 

\noindent where \( b_k^{j} \) and \( b_l^{j} \) are the components of Burger vector \( \mathbf{\textit{b}}^{j} \) along the \( \mathbf{\textit{k}} \) and \( \mathbf{\textit{l}} \) directions of the crystal axes, respectively.

\subsubsection{Irradiation-induced creep crystallography
model}

The growth of one grain under irradiation can be interrupted by the growth of a neighbouring grain. This incompatibility creates internal stresses known as creep under irradiation. Irradiation creep is intrinsically coupled with irradiation-induced growth, occurring independent of external loads.

Our analysis utilizes the crystallographic irradiation creep model for steels (~\cite{foster1972analysis},~\cite{ehrlich1981irradiation}) as implemented by \cite{patra2017crystal}, with swelling effects disregarded. This constitutive law models the creep strain rate as linearly proportional to both applied stress and radiation dose rate \( \frac{d\phi}{dt} \), expressed in terms of effective stress and strain components:

\begin{equation}
\label{TasaCreep}
\dot{\varepsilon}^{\text{creep}} = B_{0} \sigma \frac{d\phi}{dt};
\end{equation}

\noindent effective stress is denoted by \( \sigma \), \( B_0 \) is the irradiation creep compliance, and \( \frac{d\phi}{dt} \) is the radiation dose rate (in \( dpa^{-1} \)). A similar phenomenological law is assumed to represent the shear rate associated with irradiation creep at the level of slip systems, as:

\begin{equation}
\label{TasaCorteCreep}
\dot{\gamma}_{\text{creep}}^{\text{j}} = B \frac{\rho_d^j}{\rho_\text{ref}} \tau^j \frac{d\phi}{dt};
\end{equation}
 
\noindent where \( B \) is the crystallographic irradiation creep compliance, \( \tau^j \) is the resolved shear stress on slip system \( j \),  and \( \rho_\text{ref} \) is the reference line dislocation density. The crystallographic shear rate is a function of the resolved shear stress \( \tau^j \) and of the line dislocations \( \rho_d^j \) gliding on slip system \( j \). The anisotropic creep is induced by the line dislocation density \( \rho_d^j \) (weighted by \( \rho_\text{ref} \)). In Eq.~\ref{TasaCorteCreep},  such that the creep deformation (and relaxation) is higher in slip systems with greater line dislocation density. The weighting parameter \( \rho_\text{ref} \) also ensures that the crystallographic irradiation creep compliance \( B \) has the same units as the macroscopic creep compliance \( B_0 \) in Eq.~\ref{TasaCreep}.

\subsection{Meso-Scale Model}
\subsubsection{Viscoplastic Self-Consistent Model (VPSC)}
\label{VPSC}

The viscoplastic self-consistent (VPSC) polycrystalline model developed by \cite{molinari1987self}, \cite{tome1993self}, has been updated in its most recent version to include the implementation of the constitutive equations for creep and irradiation-induced growth at the grain level, as previously discussed~\cite{patra2017crystal},~\cite{tome2023material}, taking into account texture, intergranular interaction, and grain shape. 

Sliding and climbing of dislocations induce plastic shear deformation. They do not lead to volume expansion. The vector \(\boldsymbol{n}^s\) denotes the normal to the slip plane, and \(\boldsymbol{b}^s\) the slip direction (Burgers vector) on slip system \(s\) of the grain. The resolved shear stress on the slip plane and along the slip direction associated with the stress tensor \(\sigma_{ij}^c\) acting on the grain is given by:

\begin{equation}
    \tau^s = b_i^s n_j^s \sigma_{ij}^c = m_{ij}^s \sigma_{ij}^c
\end{equation}

\noindent where \( m_{ij}^s \) is the Schmid tensor, \( m_{ij}^s = \frac{1}{2} (b_i^s n_j^s + b_j^s n_i^s) \). The strain rate tensor associated with the shear rate due to irradiation creep, \(\dot{\gamma}_\text{creep}^s\) (see Eq.~\ref{TasaCorteCreep}) in system \(s\), is:

\begin{equation}
\label{TasaDeformacionCreepIrradiacion}
    \dot{\varepsilon}_{ij}^\text{(creep)}=\sum_s m_{ij}^{s} \dot{\gamma}_\text{creep}^s
\end{equation}

By incorporating the contributions from both creep and irradiation-induced growth (as described in Eqs. ~\ref{EcuacionCrecimiento} and ~\ref{TasaDeformacionCreepIrradiacion}), the resulting total strain rate in the grain is given by:

\begin{equation}
\dot{\varepsilon}_{ij}=\dot{\varepsilon}_{ij}^\text{(growth)}+\dot{\varepsilon}_{ij}^\text{(creep)}
\end{equation}

The VPSC code can be executed under different linearization schemes, and the corresponding results may differ significantly when using the same hardening parameters~\cite{tome2023material}. For the response of the effective medium (polycrystal), the tangent linearization was used in this work:

\begin{equation} \bar{\dot{\varepsilon}}_{ij}=\bar{M}_{ijkl} \bar{\sigma}_{kl}+\bar{\dot{\varepsilon}}_{ij}^0
\end{equation}

\noindent here, \(\bar{\dot{\varepsilon}}_{ij}\) and \(\bar{\dot{\sigma}}_{kl}\) represent the average (macroscopic) strain rate and stress, respectively. \(\bar{M}_{ijkl}\) denotes the macroscopic creep compliance, while \(\bar{\dot{\varepsilon}}_{ij}^0\) corresponds to the macroscopic strain rate resulting from irradiation-induced growth. These macroscopic quantities, \(\bar{M}_{ijkl}\) and \(\bar{\dot{\varepsilon}}_{ij}^0\) are initially unknown and must be determined self-consistently by enforcing that the volume-averaged stress and strain rate over all grains match those of the effective medium, namely:

\begin{equation}
    \bar{\dot{\varepsilon}}_{ij}=\langle \dot{\varepsilon}_{ij}^c \rangle;\bar{\sigma}_{kl}=\langle \sigma_{ij}^c \rangle
\end{equation}

The self-consistent polycrystalline model described above has been used by \cite{tome1996role}, \cite{turner1999self}  to model creep and irradiation-induced growth in hexagonal polycrystals, but without microstructural evolution with time. In the present work, this model is extended to include the time evolution of creep and irradiation induced growth. Like \cite{patra2017finite}, but instead of doing it in Abaqus, we use Code\_Aster. A free standardised code of nuclear interest

\subsection{Macro-Scale Model}
\subsubsection{Definition of Local Coordinate Systems}
\label{DefinicionDeSistemasLocales}

\begin{figure}[h!] 
    \centering
    \includegraphics[width=0.45\textwidth]{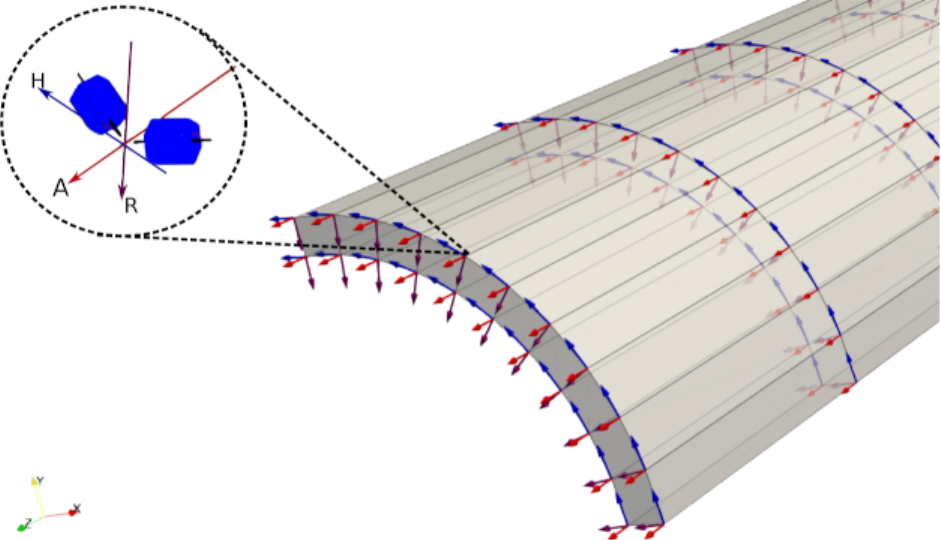}
    \caption{Relative position of the HCP axes crystals with respect to the local coordinate systems solidary to a quarter pipe. Directions A, H,
and R correspond to the axial, hoop, and radial directions of the cladding tube. 
 }
 \label{fig:SA_LA}
\end{figure}

Typically, the crystallographic texture (crystal axes, CA) and the morphological texture (ellipsoidal axes) are defined with respect to the main sample reference system (sample axes, SA). Likewise, when simulating the deformation of textured aggregates, it is common to express mechanical state variables—such as the macroscopic velocity gradient, the stress state, and internal variables—in the reference frame of the test system (test axes, TA, which are assumed to be the global axes), coincident with SA. However, when TA does not coincide with SA, it becomes necessary to define a local reference system associated with each SA at every Gauss point (see Fig. \ref{fig:SA_LA}), solve the stress or strain state in that local system, and then rotate the resulting quantities into the TA system.

Each Gauss point of the FEM mesh can be considered a polycrystal with its associated initial texture. In such case, an SA is defined at each Gauss point. If the geometry under study is curved—such as the cladding tube of the fuel element—the SA and TA systems do not coincide, since each SA rotates element by element as one moves along the circumference of the tube, as depicted in Figure~\ref{fig:SA_LA}.

This is important because the texture of these tubes is circumferential (the axis of the hexagonal prisms is mostly oriented along the hoop direction, H), see Fig. \ref{fig:ReducedTexture}(a). The orientations that carry the most relevant microstructural information for this type of texture are the Hoop face (R-A plane, since the basal pole is mostly perpendicular to the Hoop section of the tube) and the axial face (R-H plane, since the prismatic poles are mostly perpendicular to the axial section)~\cite{juarez2019evolucion}.

The Code\_Aster solver allows the definition of local coordinate systems at the Gauss points of the mesh. It is possible to manually assign the relative orientation of these local systems for each defined volume with respect to a global reference frame. This approach is manageable when working with a small number of elements; however, it becomes impractical for large meshes, especially when the geometry involves curved surfaces. In such cases, manually setting local orientations can be tedious and error-prone-particularly when analyzing contact between two curved components. Subsequently, for large meshes, a soubroutine was developed (see Appendix~\ref{apéndiceA}), using python, to automatically assign a local coordinate system to each FEM in the mesh based on its geometry. This automation significantly improves both accuracy and efficiency in defining local orientations, particularly in complex geometries and in materials exhibiting mechanical and physical anisotropic properties.

\subsubsection{VPSC-CAFEM interaction framework}
\label{Acople VPSC-FE}

\begin{figure*}[htbp]
    \centering
    \includegraphics[width=0.65\textwidth]{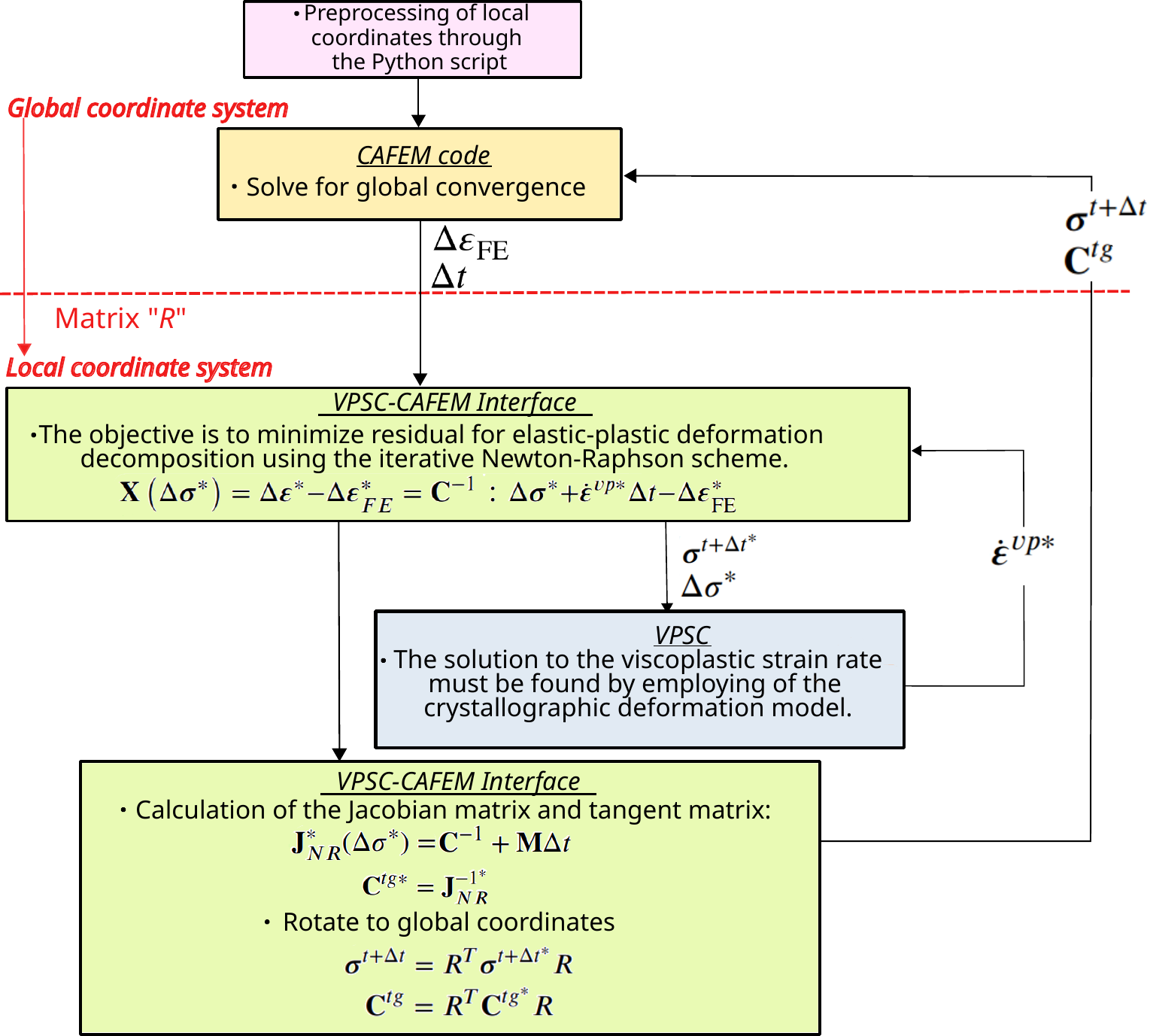}
    \caption{Flowchart of the VPSC-CAFEM interface.}
    \label{Algoritmo}  
\end{figure*}

This work uses VPSC~\ref{VPSC} to provide the constitutive deformation behavior in the FEM solver, assuming a single texture at each Gauss point in the mesh.

Existing work demonstrates VPSC integration with commercial finite element software ~\citep{segurado2012multiscale,knezevic2013integration,patra2017crystal} and fuel rod simulation tools ~\citep{liu2015vpsc}. More recently, VPSC has been successfully integrated with the open-source solver Code\_Aster~\cite{llubel2024integracion}, to development of an open-source FEM toolbox for anisotropic creep and irradiation growth in the context of nuclear-relevant materials, providing new tools for multiphysics analysis.

The implemented procedure builds upon the approach proposed by \cite{segurado2012multiscale}, incorporating a modified convergence criterion and continuous updates of the local coordinate systems. 
These material systems (defined in Section~\ref{DefinicionDeSistemasLocales}) provide the reference frame for the VPSC calculations. Subsequently, both the updated stress tensor and the consistent tangent matrix were rotated into the global coordinate system of the component, in compliance with the requirements of the Code\_Aster FEM solver. The flowchart of the VPSC-CAFEM interface is depicted in Fig.~\ref{Algoritmo}.

At this point, it is important to clarify that the standalone version of the VPSC model (hereafter referred to as VPSC-SA) relates only macroscopic stress to viscoplastic strain, without considering elastic effects. In contrast, the VPSC-CAFEM interface incorporates macroscopic elasticity from the FEM solver.

In the VPSC-CAFEM interface, we write the total strain
increment \(\Delta \boldsymbol{\varepsilon}\)  as the sum of the elastic \(\Delta \boldsymbol{\varepsilon}^{\text{e}}\) and viscoplastic \(\Delta \boldsymbol{\varepsilon}^{\text{vp}}\), i.e.:
\begin{equation}
\Delta\boldsymbol{\varepsilon} = \Delta \boldsymbol{\varepsilon}^{\text{e}} + \Delta \boldsymbol{\varepsilon}^{\text{vp}}=\textbf{C}^{-1}:\Delta\boldsymbol{\sigma} + \Delta \boldsymbol{\varepsilon}^{\text{vp}}   
\end{equation}
\noindent where \(\textbf{C}\) is the self-consistent elastic stiffness tensor of the polycrystalline material point, computed by VPSC at the beginning of each time increment, \(\Delta \boldsymbol{\sigma}\) is the Cauchy stress increment. The viscoplastic strain increment \(\Delta\boldsymbol{\varepsilon}^{\text{vp}}\) depends on the current stress state \(\boldsymbol{\sigma}\) (and not on the stress increment \(\Delta\boldsymbol{\sigma}\)), as well as on the evolution of the internal variables of the constitutive model, namely, the crystal orientations, grain morphology, and the hardening law assigned to each grain.

The VPSC-CAFEM interface is invoked by the Code\_Aster solver, providing it with an estimated strain increment \(\Delta \boldsymbol{\varepsilon}_{\text{FE}} \), a time step \(\Delta t\), and a rotation matrix \( R \) that defines the orientation of the local coordinate system relative to the global reference frame.
The VPSC-CAFEM interface rotates the stress and strain quantities to the local coordinate system, and computes the Newton-Raphson iterative scheme, that is:

\begin{equation}
\label{R1}    \Delta\boldsymbol{\varepsilon}^*=R \Delta\boldsymbol{\varepsilon} R^T
\end{equation}

\begin{equation}
\label{R2}
    \boldsymbol{\sigma}^*=R \boldsymbol{\sigma}  R^T
\end{equation}

\noindent where \(R\) is the rotation from the local coordinate system (*) to the Global coordinate system. In this work, the small strain regime is considered, which implies that the rotation matrix \(R\) remains constant over time. This is because the corotational system associated with the local coordinate frame does not undergo rotations or translations throughout the analysis.

Assuming that the total strain is accommodated elastically, the corresponding trial stress state is computed:

\begin{equation}
\label{DeltaSigmaEstrella}
    \Delta\boldsymbol{\sigma}^*=\textbf{C}: \Delta\boldsymbol{\varepsilon}^{e*}
\end{equation}

\begin{equation}
\label{SigmaEstrellat+dt}
    \boldsymbol{\sigma}^{t+\Delta t^*}=\boldsymbol{\sigma}^{t^*}+\Delta\boldsymbol{\sigma}^*
\end{equation}

\noindent given that \( \Delta\boldsymbol{\varepsilon}^{e*} = \Delta\boldsymbol{\varepsilon}^* - \Delta\boldsymbol{\varepsilon}^{vp*} \), substituting into Eq.~\ref{DeltaSigmaEstrella} and then into Eq.~\ref{SigmaEstrellat+dt}:

\begin{equation}
    \boldsymbol{\sigma}^{t+\Delta t^*}=\boldsymbol{\sigma}^{t^*}+\textbf{C}:\left(\Delta\boldsymbol{\varepsilon}^*-\Delta\boldsymbol{\varepsilon}^{vp*}     
                  \right)
\end{equation}

\noindent where the superscripts refer to the corresponding time increment. The VPSC model is called with the stress state \( \boldsymbol{\sigma}^{t+\Delta t^*} \) to compute the corresponding viscoplastic strain rate \( \dot{\boldsymbol{\varepsilon}}^{vp*} \). The VPSC-CAFEM interface also computes the difference between \(\Delta\boldsymbol{\varepsilon}^* \) and \(\Delta\boldsymbol{\varepsilon^*}_\text{FE} \), namely the residual:

\begin{equation}
    \mathbf{X}\left( \Delta\boldsymbol{\sigma}^*   \right) = \Delta\boldsymbol{\varepsilon}^*-\Delta\boldsymbol{\varepsilon}^*_{FE}=\mathbf{C}^{-1}:\Delta\boldsymbol{\sigma}^*+\boldsymbol{\dot{\varepsilon}}^{vp*}\Delta t- \Delta\boldsymbol{\varepsilon}^*_\text{FE}
\end{equation}

When the convergence criterion (given in Eq. ~\ref{ec:correcciondeTension}) is not met at iteration \(k\), the trial stress increment for the next iteration \(k+1\)  is recalculated as follows:

\begin{equation}
\label{ec:correcciondeTension}
    \left( \Delta\boldsymbol{\sigma}^*\right)_{k+1}=\left( \Delta\boldsymbol{\sigma}^*\right)_{k}-\textbf{J}^{-1^*}_{NR}((\Delta\boldsymbol{\sigma}^*)_{k}):\textbf{X}((\Delta\boldsymbol{\sigma}^*)_{k})
\end{equation}

\noindent where \(\textbf{J}_{NR}^*\) is the Jacobian of the Newton-Raphson method and is given by:

\begin{equation}
\begin{aligned}
\textbf{J}^*_{NR}(\Delta\boldsymbol{\sigma}^*) &=\frac{\partial\textbf{X}(\Delta\boldsymbol{\sigma}^*)}{\partial(\Delta\boldsymbol{\sigma}^*)}=\frac{\partial(\Delta \boldsymbol{\varepsilon}^* - \Delta \boldsymbol{\varepsilon}^*_\text{FE})}{\partial (\Delta \boldsymbol{\sigma}^* )} \\ &=\frac{\partial (\Delta \boldsymbol{\varepsilon} ^* )}{\partial (\Delta \boldsymbol{\sigma} ^* )}=\textbf{C}^{-1}+\textbf{M}\Delta t
\end{aligned}
\end{equation}

\noindent here, \(\textbf{M}\) is the viscoplastic tangent modulus calculated by VPSC in the self-consistent process.

The Code\_Aster solver computes all quantities in the global coordinate system, and therefore the VPSC-CAFEM interface rotates the stress state \(\sigma^{t+\Delta t^*}\) and the consistent tangent matrix \(\textbf{C}^{tg*} = \textbf{J}^{-1^*}_{NR}\) required by the solver to compute the strain increment for the next time step, that is:

\begin{equation}
    \boldsymbol{\sigma}^{t+\Delta t}=R^T \boldsymbol{\sigma}^{t+\Delta t^*} R   
\end{equation}

\begin{equation}
    \textbf{C}^{tg}= R^T \textbf{C}^{tg^*} R   
\end{equation}

A weighted convergence metric is employed to achieve faster convergence~\cite{mcginty2001multiscale}. The scalar convergence metric is weighted by the largest component of the strain increment, \(\Delta\varepsilon_{FE}\), as follows:

\begin{equation}
    \chi=\sqrt{\sum_i \sum_j \left(\frac{|\Delta\varepsilon_{FE}^{ij}|}{\max(|\Delta\varepsilon_{FE}^{ij}|)} \textbf{X}^{ij}  \right)^2}
\end{equation}

\noindent where \(i\) and \(j\) denote the respective tensor indices.

Validation of the VPSC-CAFEM coupling was achieved via simulations of creep and irradiation growth in a quarter-geometry cladding tube model, as detailed in Appendix~\ref{apéndiceB}.

\section{Configuration of the Numerical Simulation} \label{sec:Detalles de la simulación}

\subsection{Numerical setup} \label{Simulation setup}

The numerical study focuses on the central hexagon of the spacer grid depicted in Fig.~\ref{fig:FlexibleSpacerGrid}, selected for its geometric symmetry, which simplifies the numerical modeling process.

Fig. \ref{fig: MallaTuboyChapa}(a) depicts the cladding tube along with a section of the central hexagon of the spacer grid. The grid features dimples in contact with the tube surface, shown in Fig. \ref{fig: MallaTuboyChapa}(b), which prevent the tube from moving due to vibration induced by the coolant flow. The red dot in Fig.~\ref{fig: MallaTuboyChapa}(c) represents the point load exerted by the 45 N spring of the spacer grid, which ensures the initial contact between the tube and the dimples.

The cladding tube–grid assembly underwent uniform irradiation at \(3.6 \times 10^{-4} \, dpa \cdot hr^{-1}\), simulated over 100 increments of \(555.6 \, hr\) each. The tube’s outer surface was subjected to 15.5 MPa coolant pressure, with a concurrent 10 MPa internal pressure applied to the inner surface.
Numerous studies, including ~\cite{liu2016multiphysics}; \cite{williamson2011enhancing}; \cite{williamson2012multidimensional}, indicate that the plenum pressure on the fuel pellet during burnup can be approximated as 6–15 MPa in a first-order estimation. At the operational dose rate (3.15 dpa/year), 10 dpa and 20 dpa are reached after 3.2 and 6.4 years, respectively. The cold-worked Zircaloy-2 model was calibrated with \cite{holt1996non} irradiation growth data at 550 K. Despite no external loading, the pressure differential induces circumferential and axial stresses in the spacer grid.

\subsubsection{Crystalline texture reduction strategy} \label{sec:TexturaReducidaSeccion}

Finite element simulations that account for grain-level crystallographic behavior demand significant computational resources, especially when coupled with micromechanical approaches such as VPSC. Hence, a texture reduction strategy is employed, considering only a representative subset of crystallographic orientations is selected to reproduce the material response within a deviation margin of 10\%.

\begin{figure}[h] 
	\centering
	\includegraphics[width=0.45\textwidth]{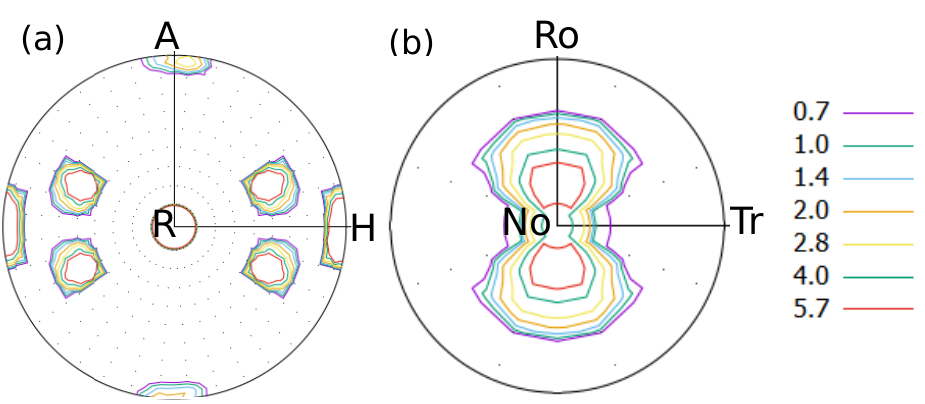}
	\caption{Basal pole figures for (a)
		reduced cladding tube texture with 7 orientations, (b) reduced dimples texture with 13 orientations. Directions A, H,
and R correspond to the axial, hoop, and radial directions of the cladding tube.
Directions Ro, Tr, and No correspond to the rolling, transverse, and normal directions of
the dimples on the spacer grid sheet.}
	\label{fig:ReducedTexture}
\end{figure}

The reduced textures proposed by \cite{patra2017finite} for both the cladding tube and the spacer grid, shown in Fig. \ref{fig:ReducedTexture}, are applied. In particular, the texture of the cold-worked Zircaloy-2 tube, originally composed of 1144 orientations, was reduced to 7 orientations. Similarly, the texture of the spacer grid was reduced from 2428 to 13 orientations. As shown in \cite{patra2017finite}, this procedure adequately captures the material anisotropy, ensuring that the Kearn factors of the reduced textures remain equivalent to those of the original ones.

The Kearn factors are defined as the average projection of the \(c\)-axes (direction cosines) of the grains that define the polycrystal, that is:
 
\begin{equation}
    F_i=\sum_g w^g (r_i^g)^2
\end{equation}

\noindent where \(w^g\) is the weight and \(r_i^g\) is the direction cosine of the \(c\)-axis along the \(i^\text{th}\) component of the \(g^\text{th}\) orientation. The use of texture reduction and the evaluation of the polycrystal response under irradiation creep conditions were discussed in detail in \cite{lhuillier1995modeling}. The Kearn factors for the tube texture are \(F_\text{H}=0.4244\), \(F_\text{R}=0.5472\), and \(F_\text{A}=0.0284\), and the corresponding values for the spacer grid are \(F_\text{Ro}=0.1907\), \(F_\text{Tr}=0.1102\), and \(F_\text{No}=0.6991\) (see Fig.~\ref{fig:ReducedTexture}).

For the correct setup of the FEM simulations, each element of the cladding tube retains the same reduced texture in the radial, hoop, and axial directions, which is then properly rotated to the laboratory axes (SA) according to the model geometry. Similarly, the texture of the spacer grid, defined along the rolling, transverse, and normal directions, is transformed to the laboratory axes in each of the corresponding elements.

\subsection{Boundary Conditions} \label{BoundaryConditions}

\begin{figure*}[h!] 
	\centering
	\includegraphics[width=0.5\textwidth]{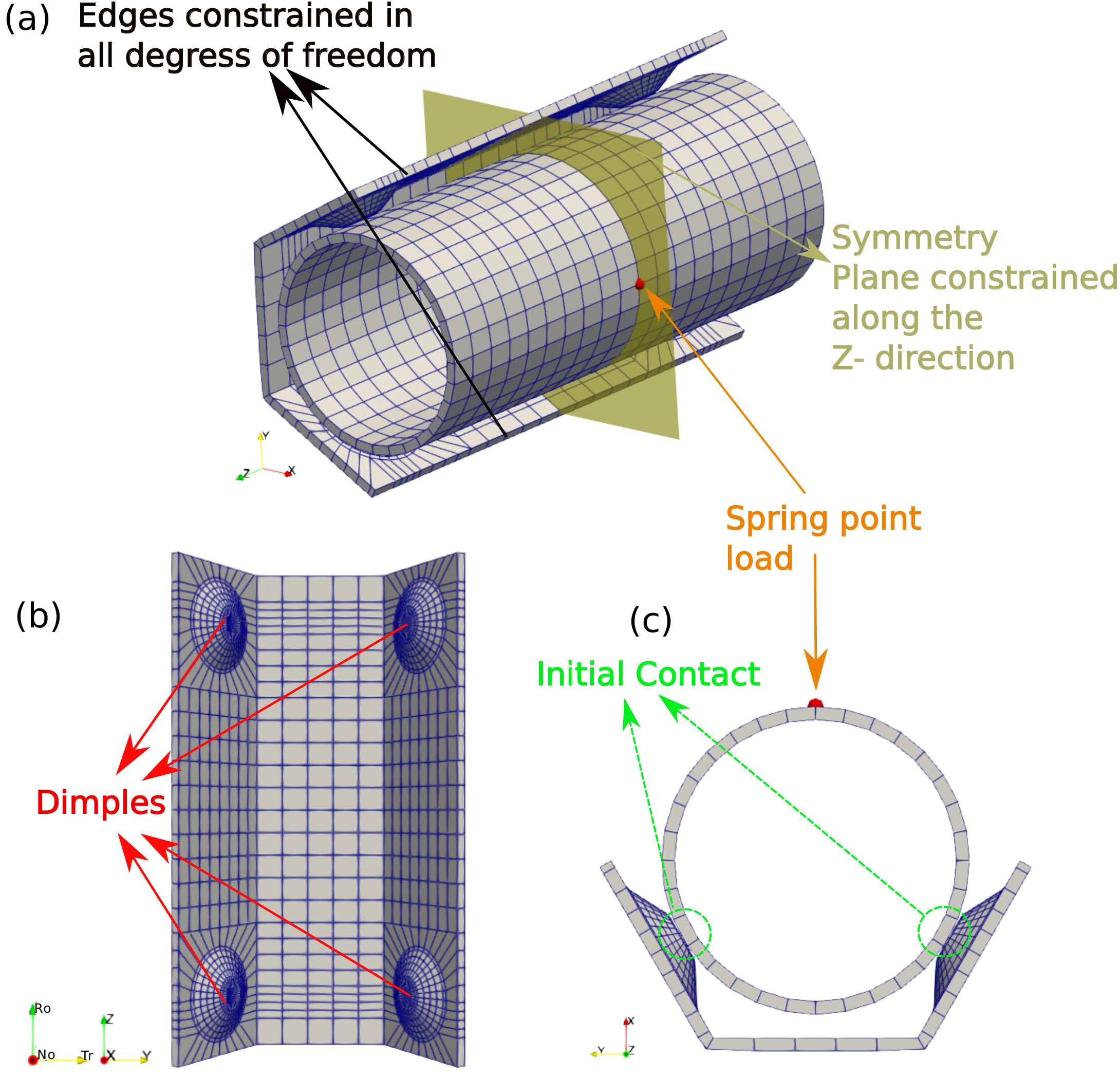}
	\caption{(a) Spreader Tube-Grid arrangement. (b) Front view of the grid with 4 support dimples. The bottom-left corner represents the material system for the grid. (c) Top view of the Tube-Grid configuration, showing the point force due to the spring.}
	\label{fig: MallaTuboyChapa}
\end{figure*}

To reduce computational cost, only half of the tube and spacer grid was modeled, taking advantage of symmetry conditions. In Fig.~\ref{fig: MallaTuboyChapa}(a), the symmetry plane used in the simulation is shown in yellow color.

The bottom edge of the cladding tube (located at the symmetry plane) is constrained in the axial direction (Z), allowing for radial expansion or contraction. Additionally, the nodes located at the discontinuous edges of the dimple-facing surfaces of the spacer grid (Fig.~\ref{fig: MallaTuboyChapa}(a)) were fully constrained in all degrees of freedom (translations and rotations) to eliminate possible rigid body motion and ensure the stability of the numerical model.

The initial configuration of the cladding tube and spacer grid was defined such that the former is held in place by the latter. To achieve this, contact was enforced during a non-irradiated time interval using a classical elastic model. Once contact was established, the simulation switched to a nonlinear contact model, initiating irradiation with the crystallographic texture properties assigned to each component (see Sec.~\ref{sec:TexturaReducidaSeccion}).

To analyze the evolution of the separation between these components during irradiation. In a nonlinear contact analysis in Code\_Aster, the state variable CLR (short for clearance) allows tracking the variation of the separation between components throughout the simulation. Additionally, the effect of the pressure conditions applied to the cladding tube is considered in this evolution.

Other approaches found in the literature regarding the tube-to-grid interaction use a classical non-penetration condition between surfaces, in which the clearance between the tube and the grid is predefined, imposing uniform displacement on all nodes of the contact regions and neglecting localized relative motion between individual nodes. A key aspect of the present work that provides a more realistic approach to the numerical application is the use of nonlinear contact implemented in Code\_Aster through master–slave surface algorithms with node-to-segment contact detection~\citep{CodeAster_U20404}, which captures differential displacements and stress distributions at the mesoscopic–macroscopic scale. The tube is defined as the master and the grid as the slave. Friction effects between both components were neglected, and a discrete formulation algorithm was selected. By removing the idealization of perfect contact and deliberately excluding friction, the model isolates the pure mechanical effects arising from the evolution of the clearance, providing a rigorous physical basis to assess potential structural failure modes.

\section{Numerical Results and Microstructural Analysis} \label{sec:resu}
\subsection{Contact Clearance Formation and Progression}
\label{Evolution of contact opening}

\begin{figure}[h!] 
	\centering
	\includegraphics[width=0.45\textwidth]{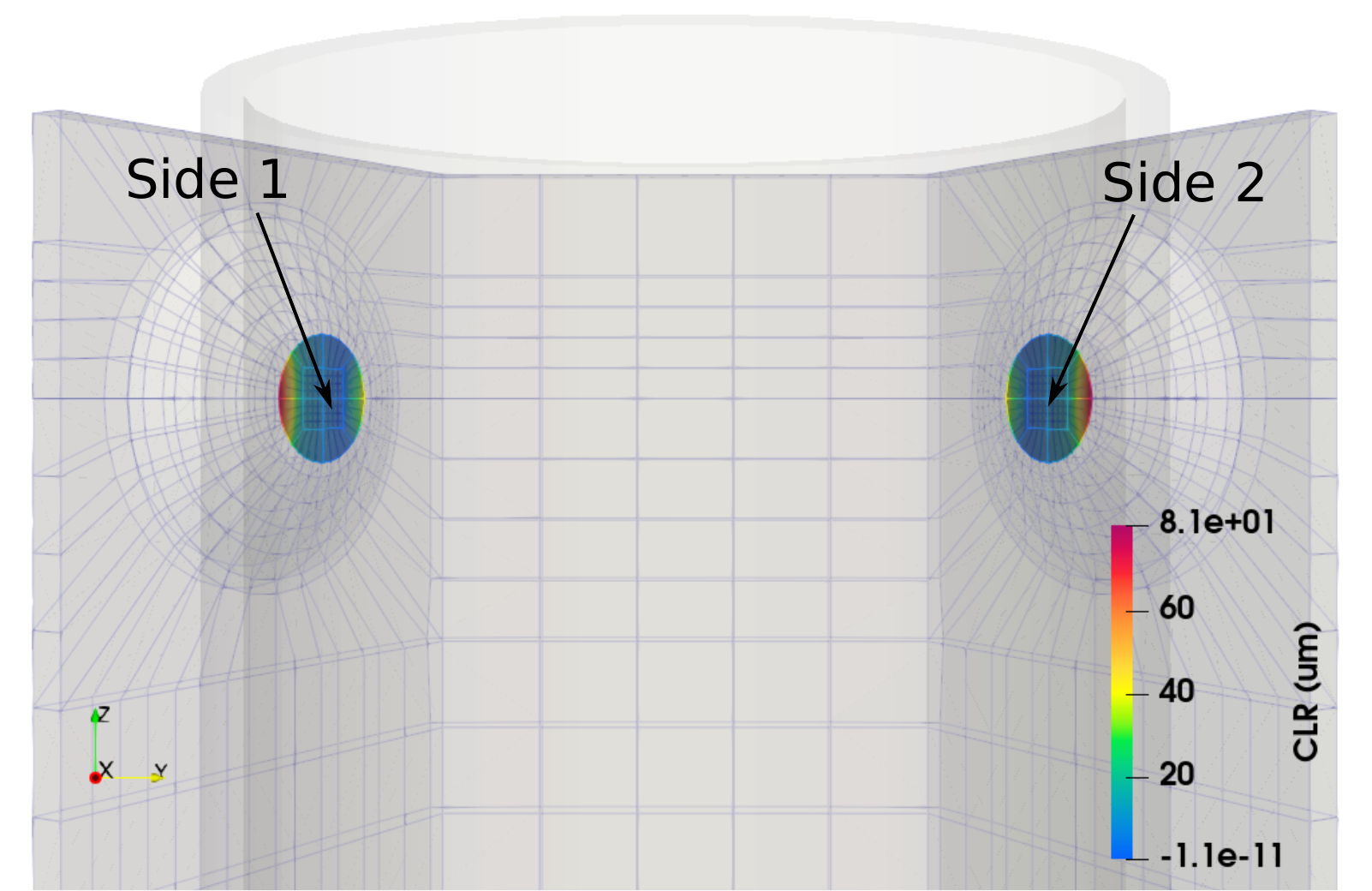}
	\caption{CLR corresponding to the slaves surfaces for non-linear contact for 20 dpa.
	}
	\label{fig:Side1andSide2}
\end{figure}

Fig.~\ref{fig:Side1andSide2} shows the contact surfaces under analysis. Considering an orientation in which the X-axis emerges perpendicularly from the plane, the contact located on the left is referred to as "Side 1", while the contact on the right is identified as "Side 2".

\begin{figure}[h!] 
	\centering
	\includegraphics[width=0.45\textwidth]{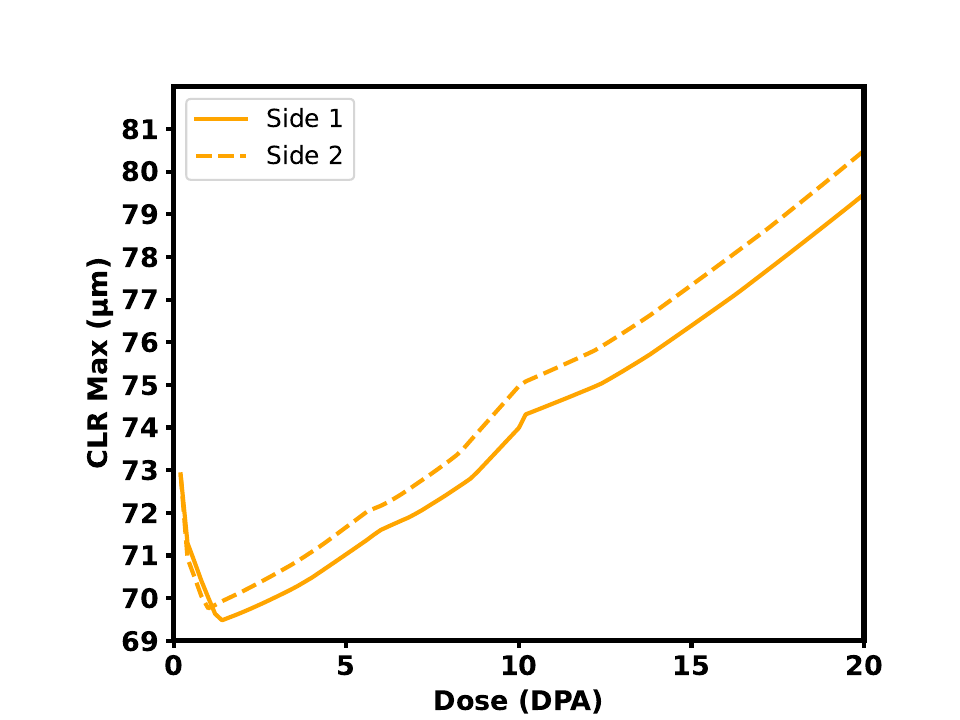}
	\caption{Maximum CLR for the texture of Test 1.
	}
	\label{fig:MaxJEU_Test1}
\end{figure}

Fig. \ref{fig:MaxJEU_Test1} shows the maximum clearance (CLR) corresponding to the textures presented in Fig. \ref{fig:ReducedTexture}. This value corresponds to the highest clearance among all nodes, specifically at the peripheral nodes of the slave surface of the dimple, where the separation from the cladding tube is greatest (see Fig.~\ref{fig:Side1andSide2}). It can be observed that, up to 1.2 dpa, the cladding tube element in contact with the dimple surface undergoes a pivoting motion around the region of highest contact (and therefore highest stress) until it fully adjusts. Beyond 1.2 dpa, both components begin to separate due to the deformation mechanisms described in Section~\ref{sec: Deformation mechanisms influencing gap opening}. Although the boundary conditions are symmetric, the slight mismatch in CLR between both sides is caused by a minimal asymmetry in the mesh, on the order of micrometers.

While in-reactor CLR data remains scarce, ex-reactor experiments on fuel rod assemblies commonly employ transverse displacement amplitudes of 5-30 µm in grid-to-rod fretting wear studies ~\cite{joulin2002effects, jiang2016grid}.
 An analytical study by~\cite{billerey2005evolution} predicted a grid-to-rod CLR of 10~µm at end-of-life for Zircaloy-4 spacer grids and cladding, accounting for irradiation growth and creep through phenomenological constitutive models. On the other hand,~\cite{patra2017finite} reported an approximate CLR of 28~µm in two of the four dimples for the grid geometry used in their study, considering the same crystallographic texture configuration and constitutive model employed in the present work. Although our simulations predict a smaller CLR of 13~µm, they confirm the occurrence of CLR opening under similar conditions and deformation mechanisms.

\subsubsection{Crystal Plasticity Effects on Clearance}
\label{sec: Deformation mechanisms influencing gap opening}

\begin{figure}[h!] 
    \centering
    \includegraphics[width=0.45\textwidth]{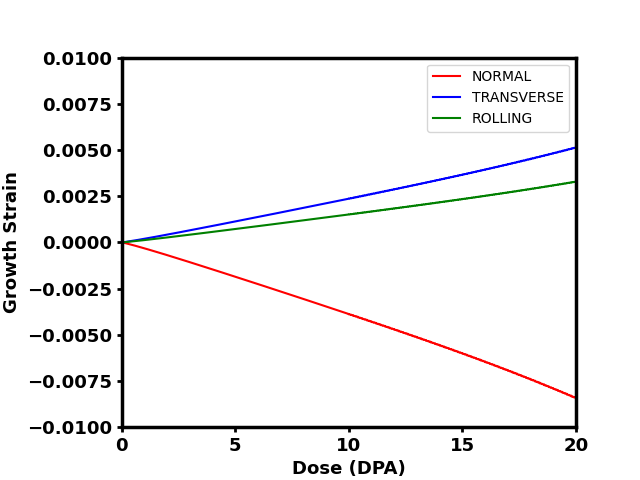}
    \caption{VPSC-SA predictions of growth strain for the grid texture in Fig. \ref{fig:ReducedTexture}(b).
 }
 \label{fig:GridGrowth-Case1}
\end{figure}

\begin{figure}[h!] 
    \centering
    \includegraphics[width=0.34\textwidth]{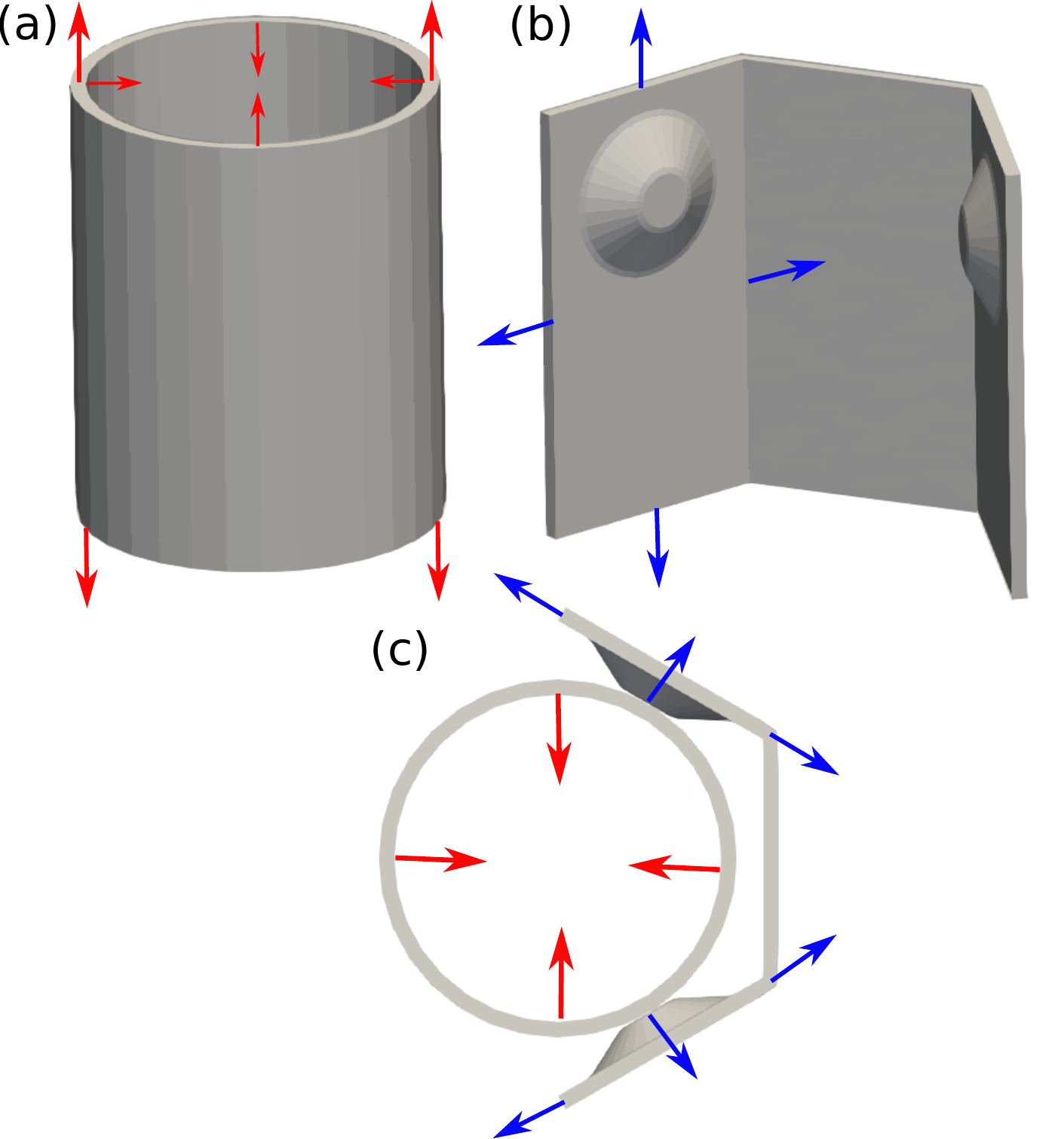}
    \caption{Schematic of texture-dependent deformation due to irradiation growth in (a): cladding tube, (b): dimple on the spacer grid, (c): top view of the cladding tube and the dimples on the spacer grid.
 }
 \label{fig:ContractionForIrradiationOnly}
\end{figure}

The cladding tube's axial elongation and radial/hoop contraction under irradiation growth (Apendix \ref{apéndiceB}, Fig. \ref{fig:GrowthVPSC-FEM}) are texture-dependent phenomena, as demonstrated by the crystallographic orientation in Fig. \ref{fig:ReducedTexture}. Both irradiation growth and creep exhibit strong microstructure and texture dependence.
As illustrated in Fig. \ref{fig:GridGrowth-Case1}, the grid texture presented in Fig. \ref{fig:ReducedTexture}(b) causes the dimples to expand along the rolling and transverse directions, while contracting along the normal direction. As shown in Fig. \ref{fig:ContractionForIrradiationOnly}(a)-(c), the basal pole-aligned direction contracts via \(c\)-axis vacancy loops, whereas interstitial loops along ⟨\(\mathbf{a}_1\), \(\mathbf{a}_2\), \(\mathbf{a}_3\)⟩ induce elongation in transverse directions—a direct consequence of the texture's anisotropy. Radiation dose accumulation drives two concurrent phenomena: (1) evolution of the \textit{CLR} (component separation) in dimples and cladding tube, and (2) relaxation of contact forces via irradiation growth and creep mechanisms (Section \ref{Evolution of contact opening}).

\subsection{Clearance Sensitivity to Texture Alignment} \label{subsec: Effect of textrue}

This section highlights the effect of texture (physically representative of the cold-work history of the material) on the CLR.

\begin{figure}[h] 
	\centering
	\includegraphics[width=0.45\textwidth]{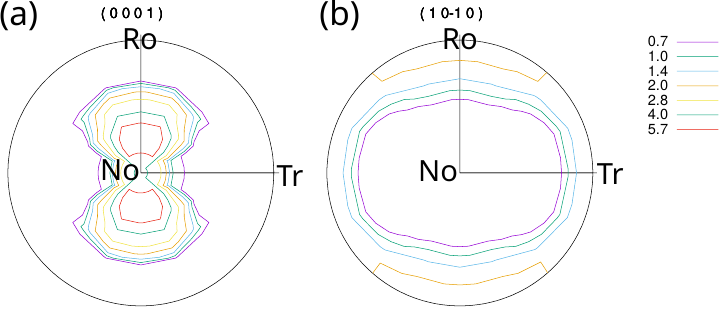}
	\caption{(a) basal pole figure for Test 1. (b)
		prismatic poles figure for Test 1.
	}
	\label{fig:TextureForTest1}
\end{figure}

\begin{figure}[h] 
	\centering
	\includegraphics[width=0.45\textwidth]{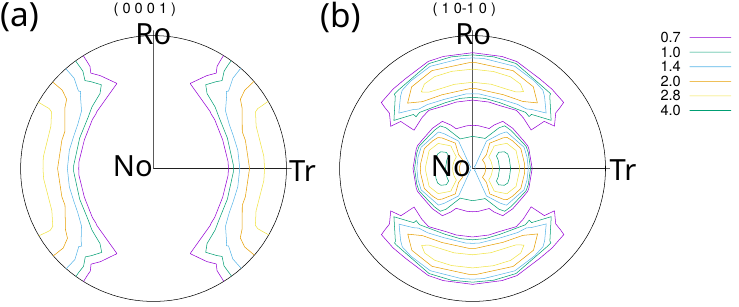}
	\caption{(a) basal pole figure for Test 2. (b)
		prismatic poles figure for Test 2. 
	}
	\label{fig:TextureForTest2}
\end{figure}

\begin{figure}[h] 
	\centering
	\includegraphics[width=0.45\textwidth]{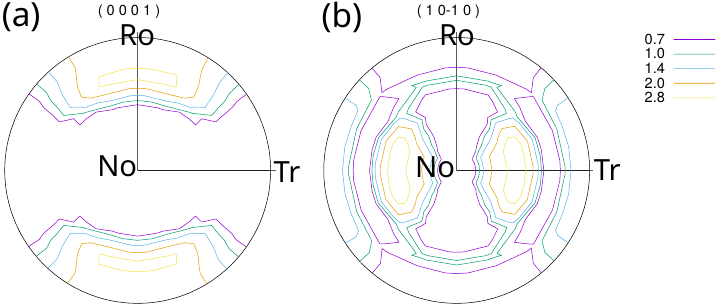}
	\caption{(a) basal pole figure for Test 3. (b)
		prismatic poles figure for Test 3. 
	}
	\label{fig:TextureForTest3}
	
\end{figure}

\begin{figure}[h] 
	\centering
	\includegraphics[width=0.45\textwidth]{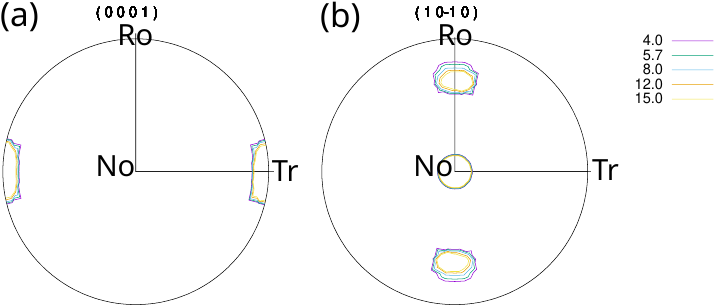}
	\caption{(a) basal pole figure for Test 4. (b)
		prismatic poles figure for Test 4. 
	}
	\label{fig:TextureForTest4}
	
\end{figure}

Four simulations were carried out in which the varying parameter was the orientation of the basal poles relative to the geometry of the dimples, while keeping the orientation of the basal poles of the cladding tube unchanged. Although these are hypothetical cases, they are useful for establishing a design parameter based on crystallographic texture. In the first test, referred to as Test 1, the basal poles are oriented along the normal direction (see Fig. \ref{fig:TextureForTest1} (a)). In the second test, referred to as Test 2, the basal poles are oriented towards the transverse direction of the dimples (see Fig. \ref{fig:TextureForTest2} (a)). In the third test, referred to as Test 3, the basal poles are oriented towards the rolling direction of the dimples (see Fig. \ref{fig:TextureForTest3} (a)). Finally, the fourth test, referred to as Test 4, corresponds to an HCP single crystal with the basal pole oriented along the transverse direction of the dimples (see Fig. \ref{fig:TextureForTest4} (a)). In Figs. \ref{fig:TextureForTest1} to \ref{fig:TextureForTest4}, directions Ro, Tr, and No correspond to the rolling, transverse, and normal directions of the spacer grid sheet.


\begin{figure}[h] 
	\centering
	\includegraphics[width=0.5\textwidth]{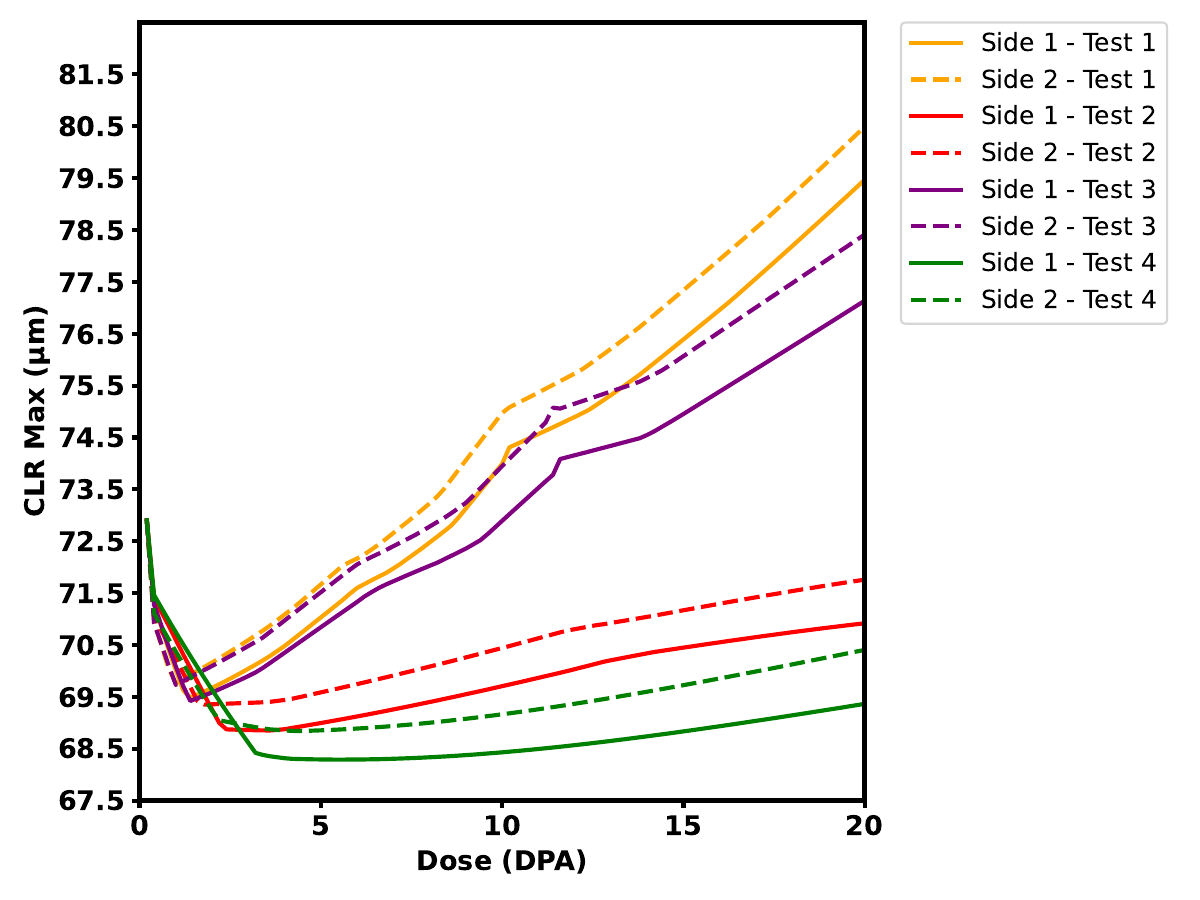}
	\caption{Comparison of the maximum CLR for all tests as a function of irradiation dose, considering the dependence on crystalline texture.
	}
	\label{fig:MaxJEUforAllTests}
\end{figure}

If the dimple sheets are processed such that the basal poles are predominantly aligned with the normal direction (Test 1), it results in a larger CLR, approximately 13~µm, which is detrimental from a fretting wear perspective (increased CLRs induce stronger flow-induced vibrations, exacerbating fretting wear damage); see the orange curves in Fig. \ref{fig:MaxJEUforAllTests}. If the basal poles are oriented along the transverse direction (Test 2), the CLR decreases significantly compared to Test 1, remaining nearly unchanged from its initial value in this Test 2; see the red curves in Fig. \ref{fig:MaxJEUforAllTests}. This behavior is associated with the higher intensity of prismatic planes oriented near the normal direction compared to the texture of Test 1 (see Fig. \ref{fig:TextureForTest2} (b)). Given the deformation mechanism under irradiation described in Section \ref{sec: Deformation mechanisms influencing gap opening}, in this test the orientation of the hexagonal \(a\)-axes lies closer to the normal direction of the grid, which results in a reduced separation between components due to expansion along this axis, as a consequence of \(c\)-axis contraction and volume conservation.
If the basal poles are oriented along the rolling direction of the grid, the CLR decreases compared to Test 1, although not as significantly as in Test 2 (see the purple curves in Fig. \ref{fig:MaxJEUforAllTests}). This is explained by a higher intensity of prismatic planes in the normal direction compared to Test 1, but lower than in Test 2 (see Fig. \ref{fig:TextureForTest3} (b)).

To strengthen the previous findings, we conducted a test referred to as Test 4, in which an HCP single crystal was used with the basal pole oriented along the transverse direction of the grid (see Fig. \ref{fig:TextureForTest4}(a)). This configuration results in the highest intensity of prismatic planes in the normal direction of the grid (see Fig. \ref{fig:TextureForTest4}(b)). In this case, the CLR was lower than in Test 2 (see green curves in Fig. \ref{fig:MaxJEUforAllTests}), confirming that a texture with basal poles oriented along the transverse direction is the most suitable for grid design.

\subsection{How Pressure Direction Reversal Modifies Clearance}

\begin{figure}[h] 
    \centering
    \includegraphics[width=0.45\textwidth]{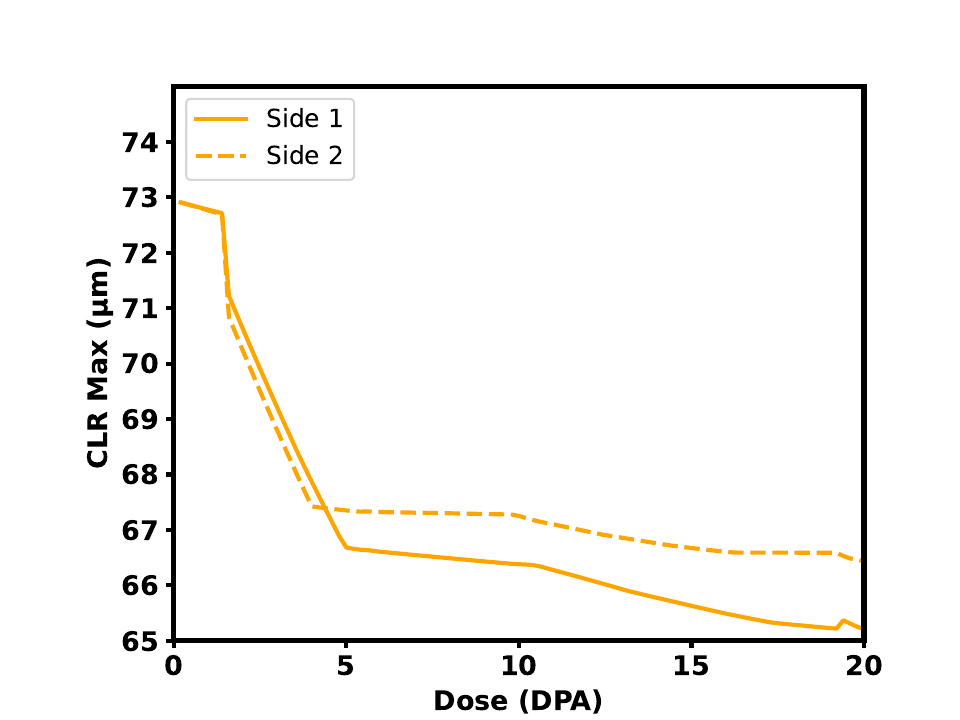}
    \caption{Maximum CLR considering effect of pressure reversal: Internal Pressure > External Pressure. 
 }
\label{fig:MaxJEUforCase1_PI}
\end{figure}

Another simulation was carried out to analyze the clearance when the pressure acting on each side of the cladding tube was inverted, changing from 15.5MPa on the outer surface and 10MPa on the inner surface to 10MPa on the outer surface and 15.5MPa on the inner surface. As a result, a net internal pressure of 5.5~MPa acts on the cladding tube, promoting creep in the hoop direction and leading to radial expansion—an accident scenario that can occur in the event of a loss-of-coolant condition. All other boundary conditions, material properties, and texture remained consistent with the simulation described in Section \ref{Simulation setup}. Fig. \ref{fig:MaxJEUforCase1_PI} shows the maximum CLR evolution for this case. It can be observed that the maximum CLR between the cladding tube and the dimples decreases as irradiation progresses. The tube's radial displacement promotes stronger spacer grid engagement, while irradiation growth and creep provide insufficient stress relaxation to offset these contact forces.

\begin{figure}[h] 
    \centering
    \includegraphics[width=0.5\textwidth]{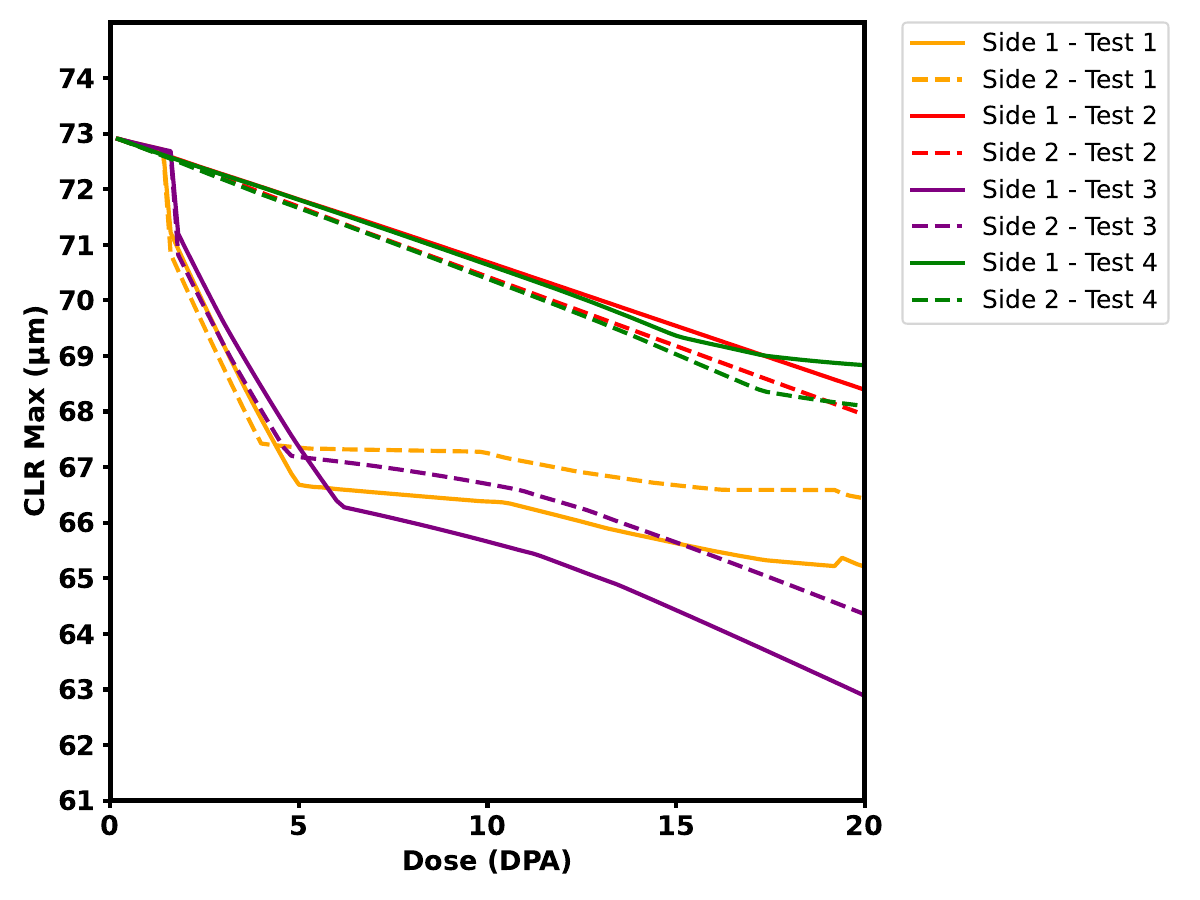}
    \caption{Comparison of the maximum CLR for all tests as a
function of irradiation dose, considering the dependence on
crystalline texture and pressure reversal. 
 }
\label{fig:MaxJEUforAllCases_PI}
\end{figure}

Considering the different texture configurations from the simulations in Section \ref{subsec: Effect of textrue}, it can be observed (see Fig. \ref{fig:MaxJEUforAllCases_PI}) that a higher presence of prismatic planes oriented near the normal direction of the dimple geometry results in greater CLR compared to cases where these planes are not predominantly aligned in this direction. However, a low presence is still preferable to a complete absence (Test 1 vs. Test 3). This is because when the \(a\)-axis of the HCP crystal is more aligned with the normal direction of the dimple geometry, growth is restricted due to the radial expansion of the tube toward the grid as a result of the tube’s net internal pressure. On the other hand, all simulations under these boundary conditions are desirable from the perspective of reducing the CLR.

Although this loading scenario is hypothetical, it confirms that CLR—even under steady-state conditions—results from the combined effects of pressure on the cladding tube and the irradiation growth and creep behavior of the material. Under reactor conditions, an increase in internal pressure may arise from the release of gas bubbles from the fuel pellets and their subsequent expansion.

\section{Conclusions} \label{sec:conclu}

In this work, the polycrystalline plasticity framework VPSC, which is capable of modeling both creep and irradiation-induced growth phenomena, was successfully integrated with the finite element solver Code\_Aster. To enable this integration, a dedicated interface was developed to facilitate the exchange of stress and strain tensors between VPSC and Code\_Aster, including the rotation of variables to a global reference frame. The additive strain decomposition in our model implementation enables analytical coupling of elasticity and viscoplasticity, overcoming access limitations to recent VPSC updates while providing an open-source solution with explicit resolution (following \cite{segurado2012multiscale,patra2017finite,knezevic2013integration}, framework).

The functionality of VPSC integrated with Code\_Aster was validated through simplified test cases, with results benchmarked against those obtained from the standalone execution of VPSC. This coupled toolbox was applied to analyze the interaction between the spacer grid and the cladding tube in a fuel rod assembly, focusing on the clearance evolution—more accurately captured through a nonlinear contact formulation. The model incorporates the actual geometry of the patented spacer grid design, including dimples, to evaluate how the separation between the grid and the fuel rod evolves as a function of irradiation

The dominant deformation mechanisms controlling clearance evolution have been identified and linked to the material's texture-dependent response. A parametric study evaluated how crystallographic texture influences dimple clearance development. The results show that when prismatic poles are aligned with the normal direction of the dimple sheet, the clearance decreases significantly, improving the wear resistance of the assembly. In contrast, when the majority of prismatic poles are aligned along the normal direction of the grid, the clearance is minimized, making this configuration more desirable. This observation is reaffirmed by the simulation of the single crystal oriented in the same way.  This behavior was consistently observed under the same irradiation dose. Finally, under reversed loading conditions—where a net outward pressure is applied—the clearance remains closed throughout the irradiation period. This finding suggests that an initial internal pressurization of the cladding may improve its performance by mitigating fretting-related degradation.

\section*{CRediT authorship contribution statement}
F. E. Aguzzi: Methodology. Software, Formal analysis, Writing.  S. M. Rabazzi: Software, Formal analysis. M. S. Armoa: Methodology, Writing. C. I Pairetti: Supervision, Writing - review \& editing. A. E. Albanesi: Supervision, Writing - review \& editing.

\section*{Declaration of competing interest}

The authors declare that they have no known competing financial interests or personal relationships that could have appeared to influence the work reported in this paper.

\section*{Acknowledgments}
The authors gratefully acknowledge the financial support from CONICET (Argentine Council for Scientific and Technical Research) and the National Agency of Scientific and Technological Promotion of Argentina (ANPCYT) for the Grant PICT-2020-SERIEA-03475. The authors also acknowledge computer time provided by CCT-Rosario Computational Center, member of the High Performance Computing National System (SNCAD, ME-Argentina).

We sincerely thank Dr. Anirban Patra (Indian Institute of Technology Bombay) for providing the texture files of the cladding tube and spacer grid, as well as for his willingness to address our inquiries. We also extend our gratitude to Dr. Martin Leonard (Institute of Physics Rosario) and Dr. Carlos Tomé (Los Alamos National Laboratory, USA) for their insightful discussions and valuable contributions to the interpretation of our results. Finally, F. E. Aguzzi gratefully acknowledges Dr. Javier Signorelli, his first mentor, with whom he took the initial steps of this work. His guidance and passion for science left an enduring mark on this research. After his passing in August 2024, the other authors joined efforts to carry forward his legacy, and their support was invaluable in bringing this work to completion. His influence will continue to resonate in the scientific community and in the hearts of those who had the privilege of knowing him.

\section*{Appendix A: Subroutine to define Local Coordinates} \renewcommand{\theequation}{A\arabic{equation}}
\renewcommand{\thesection}{A}
\refstepcounter{section} 
\label{apéndiceA}

The script makes uses of Code\_Aster utilities along with scientific libraries such as Pandas and NumPy to process structural meshes, compute the normal vectors of QUAD4 element faces (\cite{CodeAster_U30100}), and determine their nautical angles, $\theta$ and $\phi$. The following section provides a concise overview of its functionality.

\begin{figure*}[h!]
	\small
	\centering
	\includegraphics[width=0.8\textwidth]{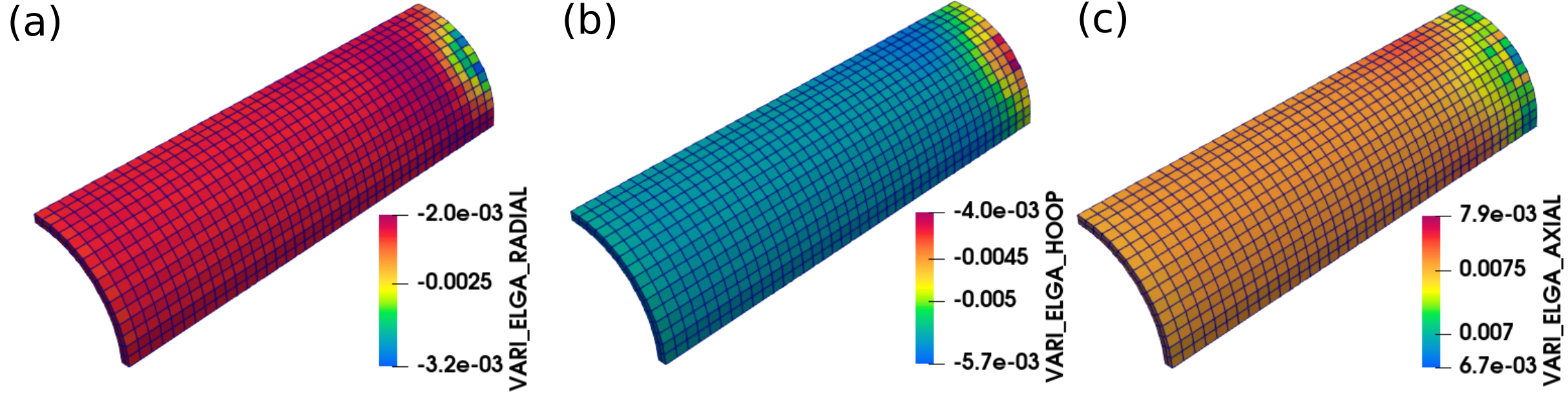}
	\caption{Distribution of (a):  Radial, (b): Hoop, and (c): Axial strains in the cladding after irradiation to 20 dpa.}
	\label{fig:DistributionOfGrowthStrain}  
\end{figure*}

\begin{itemize}
	
   \item Mesh Conversion and Reading:  
   The partition module of *Code\_Aster* is used to convert the selected mesh (MESH) into the internal MAIL\_PY format. This format enables efficient extraction of node coordinates and element connectivity. The coordinates and connectivity data are then organized into Pandas DataFrames, facilitating their manipulation.
   
   \item Extraction of QUAD4 Face Connectivities:  
   The connectivity data for faces corresponding to QUAD4 elements are filtered using the identifier (GROUP\_Q4). This ensures that only the relevant faces are processed for the analysis.
   
   \item Computation of Normal Vectors:  
   For each QUAD4 face, the coordinates of its four nodes, ordered counterclockwise, are identified. From the first node, two vectors are constructed within the plane of the face: one toward the second node and another toward the third. This guarantees that both vectors lie in the same plane. The cross product of these vectors is then computed and normalized, yielding the normal vector ``\(n\)'' to the face.
   
   \item Determination of Nautical Angles:  
   The computed normal vectors are transformed into angular coordinates by calculating the inclination (\(\theta\)) and azimuth (\(\phi\)) based on the three components  
   \([n_x,n_y,n_z]\) of the normal vector  
   \(n\):

\begin{align*}
\theta = \arcsin{\frac{n_z}{r}}, \quad \phi= \arctan2(n_x,n_y)
\end{align*}

where \(r=\sqrt{n_x^2+n_y^2+n_z^2}\)  
is the magnitude of the normal vector. The values are converted to degrees and rounded to two decimal places.

\item Association with HEX8 Volumes:  
The angles obtained for QUAD4 faces are linked to the corresponding HEX8 volumes in the mesh, filtered based on the target group (GROUP\_HEX8).

\item Script Output:  
The output is a table containing the HEXA8 volume identifier and the nautical angles (\(\phi,\theta\)) corresponding to the selected QUAD4 face. This table is subsequently read by *Code\_Aster* to assign the new local coordinate system to the volumetric element using the \texttt{AFFE\_CARA\_ELEM} command with the \texttt{MASSIF} option.

\end{itemize}

The use of this script enables the automation of geometric analysis in finite element simulations, significantly reducing manual effort and increasing process accuracy. This type of tool is particularly valuable in scenarios requiring detailed integration between polycrystalline models and finite element software, such as in the development and implementation of user material subroutines.

In this context, the definition of local coordinate systems plays a crucial role, as it allows Code\_Aster, when calling the user material subroutine, to provide a rotation matrix  
\(R\) (see Equations \ref{R1} and \ref{R2}), which defines the transformation from the local reference system to the global system (TA). This capability not only ensures the correct interpretation of orientations but also facilitates more accurate and representative simulations of the material's anisotropic behavior. The subroutines used to assign material coordinate systems in Code\_Aster is available at \citep{RotaSanti}.

\begin{figure}[h!]
	\small
	\centering
	\includegraphics[width=0.45\textwidth]{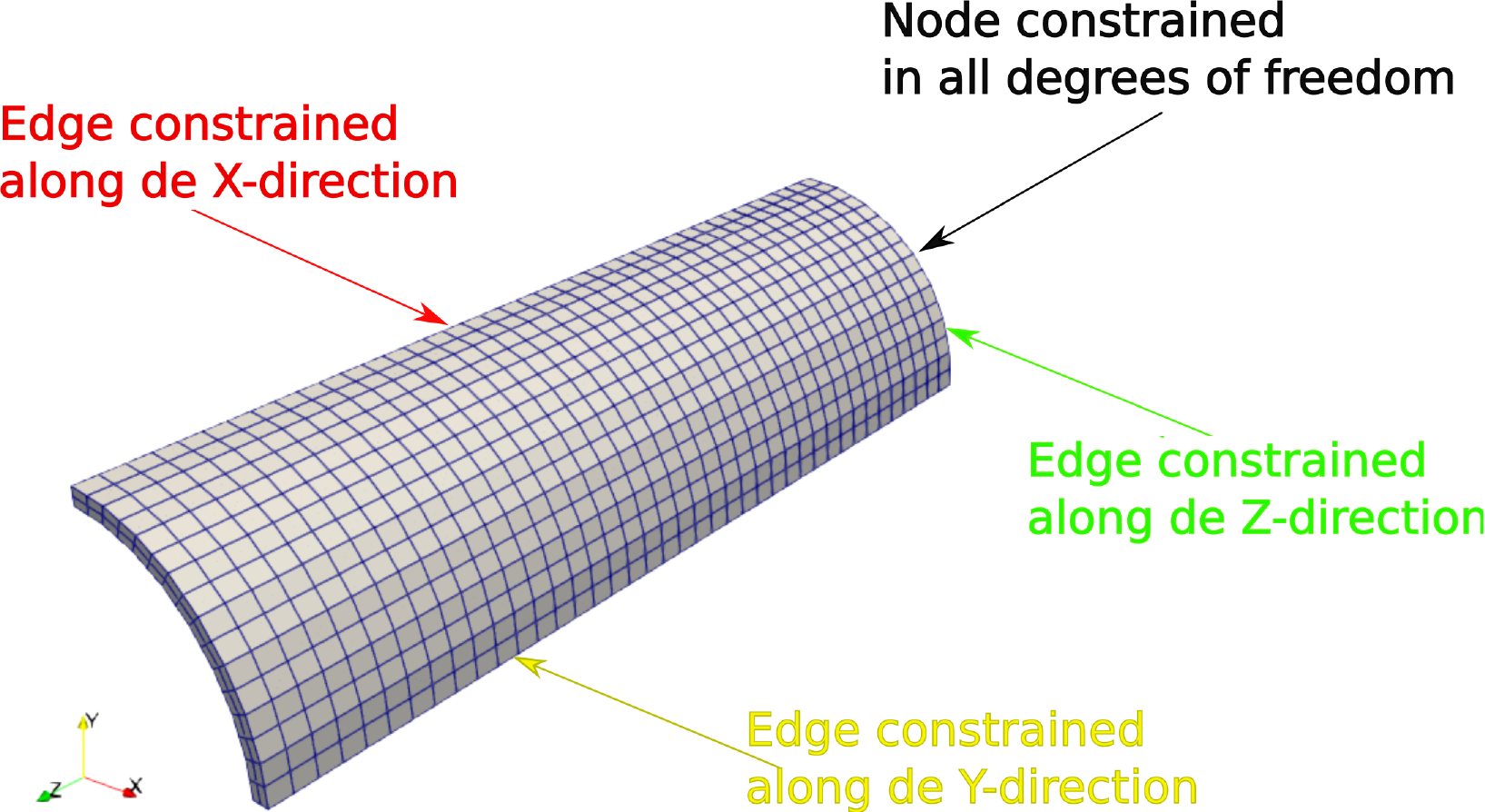}
	\caption{Mesh and boundary conditions for quarter geometry of the cladding tube.}
	\label{MallayCondicionesdeBordeTubo}  
\end{figure}

\section*{Appendix B: Validation Test} 
\renewcommand{\theequation}{B\arabic{equation}}
\renewcommand{\thesection}{B}
\refstepcounter{section} 
\label{apéndiceB}

Some simple simulations were carried out to validate the integration of the VPSC-CAFEM interface. They were applied to a geometry corresponding to a quarter tube, which would emulate the cladding tube. In this case, the reduced texture (\ref{sec:TexturaReducidaSeccion}) still characterises this behaviour. The cladding tube (see Fig. \ref{MallayCondicionesdeBordeTubo}) has an outer diameter of \(11.094~mm\) and an inner diameter of \(10.134~mm\), with a length of \(27.2~mm\). The tube within the spacer grid constitutes the fuel element. The geometry was meshed with 160 hexahedral elements, each with 8 Gauss points—20 axial elements and 8 circumferential elements.  

As mechanical boundary conditions, axisymmetric constraints were applied such that the edges parallel to the axial axis of the tube are restricted in the \(X\) and \(Y\) directions, respectively. It is hypothesized that one of the extremities of the tube may correspond to the base. Displacement in the axial direction is constrained. While the other end is permitted this degree of freedom, this would demonstrate growth by irradiation in this direction. Two nodes have been identified as the primary sites of restriction on all degrees of freedom. These are located in the medial aspect of the base. These conditions preclude the possibility of rotations (see Fig. \ref{MallayCondicionesdeBordeTubo}).

The material parameters used were taken from \cite{patra2017crystal} and are shown in Table \ref{tab:zircaloy_params}.  The tube was irradiated at a rate of \(3.6 \times 10^{-4}~dpa \cdot hr^{-1}\) over 480 time steps of \(111.12~hr\) each, with an internal pressure sufficiently small to run the simulation while assuming zero external load. In this way, it is possible to isolate and observe the phenomenon of irradiation-induced growth only.

Following the simulation of a cumulative irradiation of 20 dpa, the spatial distribution of deformation due to growth under irradiation is evident in Fig. \ref{fig:DistributionOfGrowthStrain} (a)–(c). Is composed of three distinct elements, designated as Radial, Hoop, and Axial, respectively, with the cladding tube serving as the underlying reference point.

In contrast, a uniform distribution of deformation is evident along the circumferential direction. This phenomenon transpires as a consequence of symmetrical boundary conditions, in conjunction with the observation that the sole agent of action is irradiation, with the absence of external loads. As a result of the tube texture, positive growth occurs in the Axial direction under the influence of irradiation. Pursuant to the principle of volume conservation, the deformations in the Radial and Hoop directions are negative.

\begin{figure}[h!]
	\small
	\centering
	\includegraphics[width=0.45\textwidth]{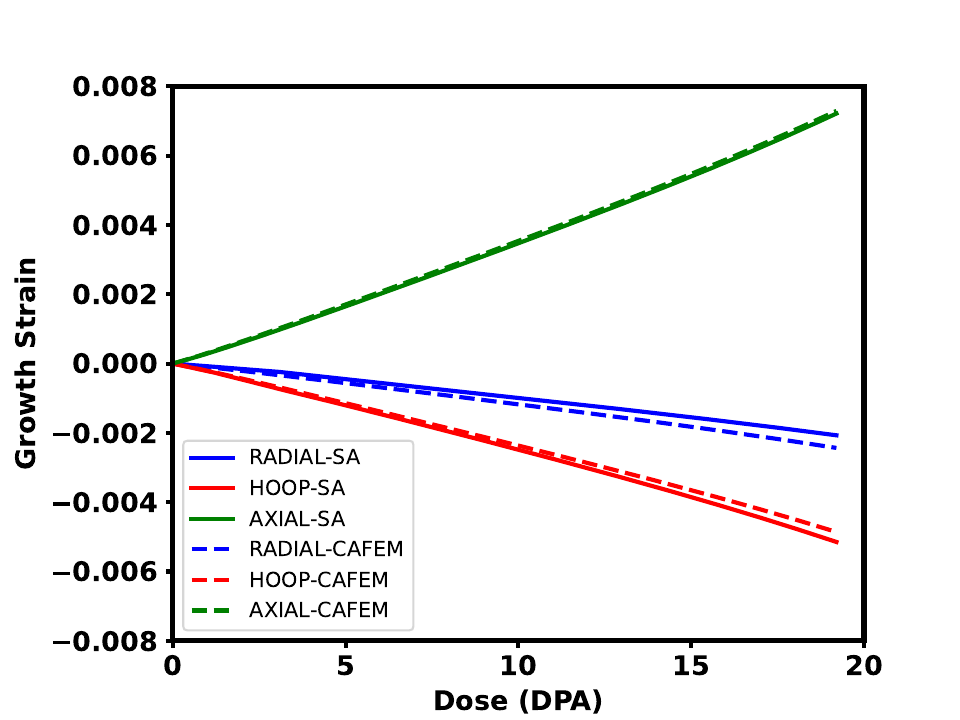}
	\caption{Evolution of growth strains (in the absence of applied stress). }
	\label{fig:GrowthVPSC-FEM}  
\end{figure}

A Gauss point was selected at the midpoint of the tube to validate that the outcomes yielded by VPSC-CAFEM are consistent with those predicted by VPSC-SA (see Fig. \ref{fig:GrowthVPSC-FEM}. This figure shows the evolution of Radial, Hoop, and Axial strains as a function of irradiation dose. It is observed that both predictions (VPSC-CAFEM and VPSC-SA) show good agreement.

\begin{figure}[h!]
	\small
	\centering
	\includegraphics[width=0.45\textwidth]{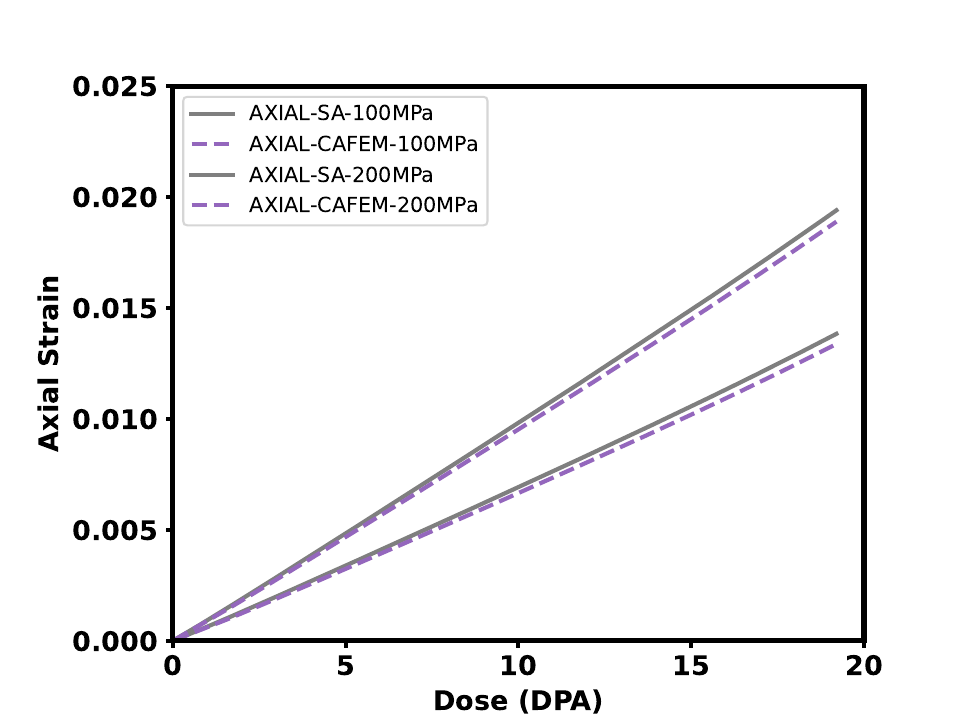}
	\caption{Axial creep strain under irradiation for  100 MPa and 200 MPa.}
	\label{fig:CreepVPSC-FEM}  
\end{figure}

To verify the prediction of irradiation creep, a similar comparison is carried out between the previously discussed VPSC-CAFEM and VPSC-SA predictions, this time with the application of different axial loads. Fig. \ref{fig:CreepVPSC-FEM}  shows the comparison of Axial strain between VPSC-CAFEM (at the same Gauss point as in the previous case) and VPSC-SA for axial loads of 100~MPa and 200~MPa, respectively. Since the strain component due to irradiation-induced growth does not depend on the applied stress, the resulting strain after applying an axial load will be attributed to the irradiation creep component.


\begin{table*}[h]
\small
    \centering
    \caption{Model parameters for Zircaloy-2}
    \renewcommand{\arraystretch}{1.2}
    \begin{tabular}{lp{6cm}p{8cm}}  
        \toprule
        \textbf{Parameter} & \textbf{Value} & \textbf{Meaning} \\
        \midrule
        $C_{11}, C_{22}, C_{33}$ & 143.5, 143.5, 164.9 & \multirow{2}{=}{Elastic constants in GPa (assumed to be the same as that for pure Zr). Values taken from \cite{simmons1965single} and \cite{kocks2000texture}.} \\
        $C_{12}, C_{13}, C_{23}$ & 72.5, 65.4, 65.4 & \\
        $C_{44}, C_{55}, C_{66}$ & 32.1, 32.1, 35.5 & \\
        \midrule
        $f_r$ & 0.97 & Fraction of point defects that recombine during the cascade \\
        $f_{ic}$ & 0.13 & Fraction of interstitials that form clusters \\
        $B$ & $5.0 \times 10^{-5}$ MPa$\cdot$dpa$^{-1}$ & Crystallographic irradiation creep compliance \\
        $p_{ref}$ & $2.26 \times 10^{14}$ m$^{-2}$ & Weighting factor for line dislocation density in irradiation creep model \\
        $b^i, \ j = \alpha_1, \alpha_2, \alpha_3$ & $3.0 \times 10^{-10}$ m & Magnitude of Burgers vector along the prismatic directions \\
        $b^i, \ j = c$ & $5.0 \times 10^{-10}$ m & Magnitude of Burgers vector along the basal direction \\
        \bottomrule
    \end{tabular}
    \label{tab:zircaloy_params}
\end{table*}
\FloatBarrier
\printcredits

\bibliographystyle{cas-model2-names}

\end{document}